\newif\ifsubmode
\submodefalse

\newif\ifprintfig
\printfigtrue

\newif\ifemulate
\emulatetrue

\ifsubmode
  \documentclass[12pt,preprint]{aastex}
  \received{}
  \accepted{}
  \journalid{}{}
  \articleid{}{}
\else
   \documentclass{emulateapj}
   \submitted{{\it Submitted for publication in AJ}}
\fi

\renewcommand{\thefootnote}{\arabic{footnote}}

\def\lesssim{\mathrel{\hbox{\rlap{\hbox{\lower4pt\hbox{$\sim$}}}\hbox{$<$}}}}
\def\gtrsim{\mathrel{\hbox{\rlap{\hbox{\lower4pt\hbox{$\sim$}}}\hbox{$>$}}}}

\usepackage{hyperref}
\usepackage{pdfpages}
\usepackage{color}

\def\spose#1{\hbox to 0pt{#1\hss}}
\def\simlt{\mathrel{\spose{\lower 3pt\hbox{$\mathchar"218$}}
     \raise 2.0pt\hbox{$\mathchar"13C$}}}
\def\simgt{\mathrel{\spose{\lower 3pt\hbox{$\mathchar"218$}}
     \raise 2.0pt\hbox{$\mathchar"13E$}}}

\slugcomment{Draft 7/2/15}

\shorttitle{Dwarf Galaxy Searches with RRL}
\shortauthors{Baker \& Willman}

\begin{document}

\title{Charting Unexplored Dwarf Galaxy Territory With RR Lyrae}

\author{Mariah Baker\altaffilmark{1}, Beth Willman\altaffilmark{1}}

\altaffiltext{1}{Department of Astronomy, Haverford College, Haverford, PA 19041, USA, beth.willman@gmail.com}

%%%%%%%%%%%%%%%
% Start the abstract on a fresh page
%%%%%%%%%%%%%%%

\ifsubmode\else
  \ifemulate\else
     \clearpage
  \fi
\fi

%%%%%%%%%%%%%%%
% Use a small baselineskip, unless in submission mode.
%%%%%%%%%%%%%%%

\ifsubmode\else
  \ifemulate\else
     \baselineskip=14pt
  \fi
\fi

\begin{abstract} 

Observational biases against finding Milky Way (MW) dwarf galaxies at low Galactic latitudes (b $\lesssim$ 20$^{\circ}$) and at low surface brightnesses ($\mu_{\rm V,0} \gtrsim$ 29 mag arcsec$^{\rm -2}$) currently limit our understanding of the faintest limits of the galaxy luminosity function.  This paper is a proof-of-concept that groups of two or more RR Lyrae stars could reveal MW dwarf galaxies at d $>$ 50 kpc in these unmined regions of parameter space, with only modest contamination from interloper groups when large halo structures are excluded.   For example, a friends-of-friends (FOF) search with a 2D linking length of 500 pc could reveal dwarf galaxies more luminous than M$_{\rm V}$ = -3.2 mag and with surface brightnesses as faint as 31 mag arcsec$^{\rm -2}$ (or even fainter, depending on RR Lyrae specific frequency).  Although existing public RR Lyrae catalogs are highly incomplete at d $>$ 50 kpc and/or include $<$1\% of the MW halo's volume, a FOF search reveals two known dwarfs (Bo\"otes I and Sextans) and two dwarf candidate groups possibly worthy of follow-up.  PanSTARRS 1 (PS1) may catalog RR Lyrae to 100 kpc (in the absence of Galactic extinction) which would include up to $\sim$15\% of predicted MW dwarf galaxies.  Groups of PS1 RR Lyrae should therefore reveal very low surface brightness and low Galactic latitude dwarfs within its footprint, if they exist.  With sensitivity to RR Lyrae to d $\gtrsim$ 600 kpc, LSST is the only planned survey that will be both wide-field and deep enough to use RR Lyrae to definitively measure the Milky Way's dwarf galaxy census to extremely low surface brightnesses, and through the Galactic plane.

\end{abstract}


\section{Introduction}\label{intro_sec}
\renewcommand{\thefootnote}{\arabic{footnote}} 

The Milky Way's (MW's) dwarf galaxy population provides a unique laboratory to study galaxies at the bottom of the galaxy luminosity function, and to quantify tensions between cosmological predictions and observations on sub-galactic scales.  For example, the discrepancy between the thousands of dwarf galaxy-mass dark matter halos predicted to orbit the Milky Way \citep[e.g.][]{Springel08a,ELVIS14a} and the $\sim$35 dark matter dominated dwarf galaxies observed to orbit the MW \citep{McConnachie12,Bechtol15a,Koposov15a} has been long referred to as the ``missing satellite problem'' \citep{klypin99a,moore99a}.  The discrepancy between the predicted and observed circular velocities of the most massive MW dwarf satellites has been coined the ``too big to fail" problem \citep{boylankolchin11a,boylankolchin12a}.  Most recently, increasing attention has been paid to possible differences between the predicted and observed spatial distributions of dwarf galaxies around the MW, M31 and other MW-analog galaxies \citep[e.g.][]{Kroupa05a,Pawlowski12a,Ibata14a,Phillips15a}.  Resolving any of these open puzzles about the MW's dwarf satellites, or measuring the empirical relationships needed to understand galaxy formation at its lower limit (for example, the empirical size-luminosity-distance relationship), requires a more complete and unbiased census of MW dwarfs.  

\begin{figure*}
\plottwo{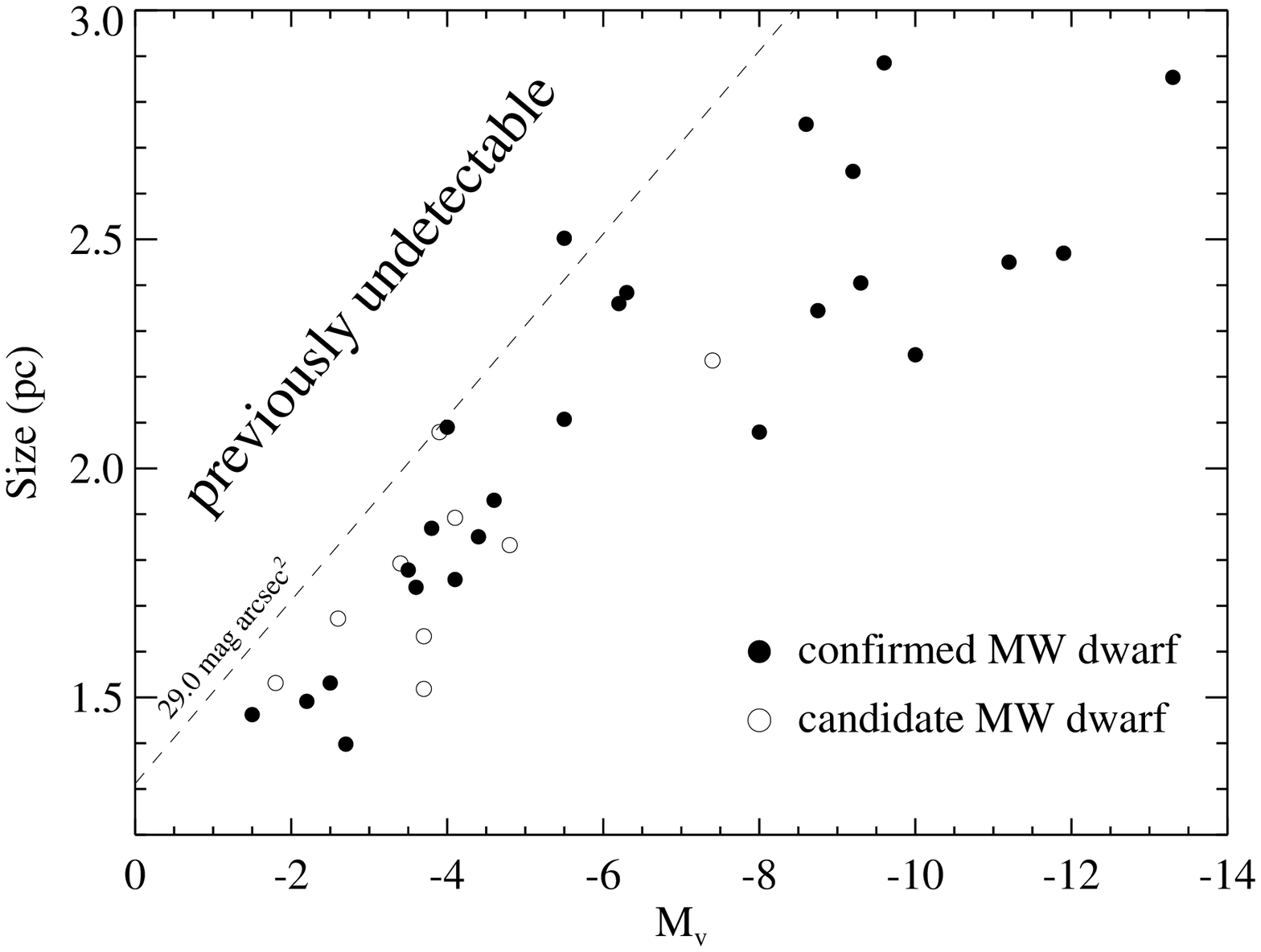}{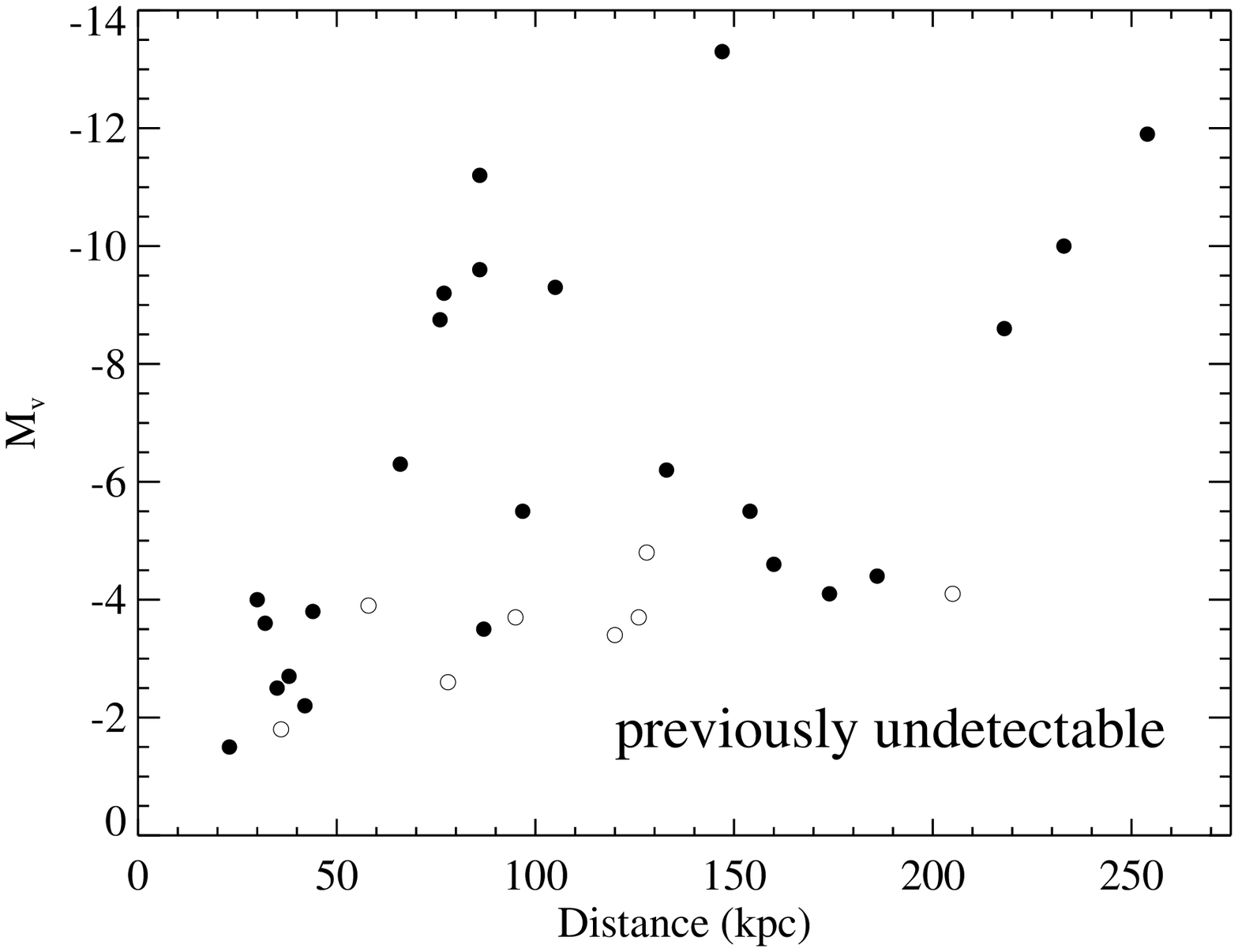}
\caption{The observed size-luminosity-distance relations of confirmed (closed points) and recently published candidate (open points) Milky Way dwarf galaxies. A line of constant central V-band surface brightness is overplotted on the Left panel.    Observational bias against finding very low surface brightness dwarfs (upper left portion of the left-hand panel) and distant, faint dwarfs (bottom right portion of the right-hand panel) continue to limit our census of Milky Way dwarfs.  This figure includes 33 MW satellites: Hercules, Boo I, Boo II, Leo IV, Leo V, Pisces II, CVn I, CVn II, UMa I, UMa II,  Coma Berenices, Will 1, Seg 1, Seg 2, Draco, Fornax, Sculptor, Carina, Sextans, Leo I, Leo II, Ursa Minor \citep[][and references therein]{Sand12a}, Leo T \citep{deJong08a}, Ret II, Eri 2, Tuc 2, Hor I, Pic 1, Phe 2 \citep{Bechtol15a}, Gru 1 \citep{Koposov15a}, Hor 2 \citep{Kim15d}, Peg 3 \citep{Kim15c}, Triangulum 3 \citep{Laevens15a}, Hydra 2 \citep{Martin15a}.}
\label{demographics}
\end{figure*}

The current detection biases against low surface brightness and low Galactic latitude dwarfs create an incomplete picture of the MW's dwarf galaxy population.
 For example, some theories predict the existence of extremely low surface brightness ``stealth'' \citep{bullock10a} and ``fossil" galaxies \citep{Bovill11a} that would have evaded detection with existing techniques. Figure~\ref{demographics} shows that no dwarfs are known to exist in the larger size - lower luminosity space that such extreme dwarf galaxies may occupy.  This demographic gap may reflect something physical about the formation of low luminosity dwarf galaxies, or may simply reflect the detection bias against dwarfs with V-band central surface brightnesses fainter than $\sim$29 mag arcsec$^{-2}$ \citep[e.g][]{Koposov08a,Walsh09,Bechtol15a}.  Moreover, MW dwarf galaxies are detected and studied by their resolved stellar populations.  The increasing number of foreground stars at Galactic latitudes lower than $b \sim$ 20$^{\circ}$ therefore limits the detectability of dwarf galaxies over a significant fraction of the MW's volume \citep{Walsh09}. 

RR Lyrae (RRL) stars may provide a new means to detect previously unseen MW dwarf galaxies.  RR Lyrae are short-period variable stars found in old and metal-poor stellar populations.  At least one RR Lyrae has been found in each Milky Way dwarf galaxy searched for such stars (see Section~\ref{sec_dwarfRRL}). There are two main categories of RR Lyrae stars: type $c$ RR Lyrae stars (RRc), which pulsate in the first overtone and have periods between 0.25 and 0.4 days, and type $ab$ RR Lyrae stars (RRab), which pulsate in the fundamental mode and have periods between 0.4 and 0.8 days \citep{Smith95, Vivas04a, Sesar07}. RR Lyrae have light curves that exhibit distinct shapes, periods and amplitudes, making them relatively easy to identify in time domain surveys.  RRab stars are also particularly good standard candles, with a well calibrated luminosity-[Fe/H] relation at optical magnitudes \citep{Chaboyer99}:

\begin{equation}
M_{V} = (0.23 \pm 0.04) [Fe/H] + (0.93 \pm 0.12) 
\end{equation}

These combined properties make RR Lyrae stars valuable tracers of structure throughout the Milky Way's halo.  Until now, RR Lyrae have primarily been used to map streams and unbound overdensities in the Milky Way's halo \citep[e.g.][]{Vivas06a,Watkins09a,Drake13a,Sesar13a,Duffau14a,Simion14a}.  However, as wide-field time domain surveys become sensitive to RR Lyrae at increasing distances into the MW's halo \citep[e.g.][]{Oluseyi12a}, RR Lyrae could be used as discovery tools for dwarf galaxies themselves (see Section~\ref{sec_data}).  \citet{Sesar14} proposed that single RR Lyrae in the outer halo could be beacons of underlying, extremely low luminosity (L $<$ 1000 L$_{\odot}$) dwarf galaxies that are otherwise invisible.  They demonstrate that stacked star catalogs around single, distant RR Lyrae stars provide interesting constraints on the presence of extremely low luminosity galaxies around the Milky Way. Another type of variable star, Cepheids, have already been used to identify a candidate dwarf galaxy near the Galactic plane \citep{Chakrabarti15a}.

In this paper, we investigate RR Lyrae stars in wide-field, time domain surveys as a possible tool for the discovery of dwarf galaxies with L $>$ 1000 L$_{\odot}$.  We are most interested in RRL as possible tracers of unseen dwarfs at low Galactic latitudes and/or with V-band central surface brightnesses fainter than 29 mag arcsec$^{-2}$.  In Section~\ref{sec_data}, we review time domain surveys that are sensitive to RR Lyrae stars at halo distances, and use the ELVIS simulations \citep{GarrisonKimmel14a} to quantify the fraction of predicted dark matter sub-halos within the grasp of planned surveys.  In Section~\ref{sec_sim}, we use the observed RR Lyrae populations of Milky Way dwarf galaxies to simulate reasonable representations of RR Lyrae in dwarf galaxies with a range of sizes and luminosities, apply a friends-of-friends algorithm to simulated dwarfs to assess their detectability as groups of RR Lyrae stars, and apply a friends-of-friends algorithm to simulated stellar halos to assess the extent of interloper groups of field RR Lyrae stars. We summarize the detectability of our simulated dwarfs as groups of RR Lyrae stars in Section~\ref{sec_detect} and present the results of a friends-of-friends group finder applied to existing RR Lyrae catalogs in Section~\ref{sec_ObsGroups}.

\section{Time Domain Surveys Sensitive to Halo RR Lyrae}\label{sec_data}

\subsection{Summary of Survey Properties}

Numerous time domain surveys with published RR Lyrae catalogs are sensitive to RR Lyrae in the Milky Way's halo.  Existing RR Lyrae catalogs that cover more than 1\% ($\sim$400 deg$^{2}$) of the sky at halo distances include: LINEAR \citep{Sesar11,palaversa13}, PanSTARRS 1 - SDSS \citep{Abbas14}, La Silla QUEST \citep[LSQ,][]{Zinn14a}, SEKBO \citep{Keller08a}, the three component Catalina Surveys\footnote{Including the Catalina Schmidt Survey \citep[CSS,][]{Drake13a}, the Sliding Spring Survey \citep[SSS,][]{torrealba15a}, and the Mount Lemmon Survey \citep[MLS,][]{Drake13b}} \citep{Drake09a}, and Stripe 82 \citep{Sesar10a}.  (Although SDSS Stripe 82 only covers 270 deg$^2$ of sky, we include it in this list because of its excellent depth and RR Lyrae completeness.) 

Ongoing and future surveys that will yield catalogs of RR Lyrae stars at halo distances include: the  VISTA Variables in the Via Lactea \citep[VVV,][]{Saito12,gran15a}, Palomar Transient Factory \citep[PTF, and its successors Intermediate Palomar Transient Factory and Zwicky Transient Facility,][J. Surace private communication 2015]{Law09a,Bellm14}, PanSTARRS 1 (B. Sesar private communication 2015, N. Hernitschek et al in preparation), and LSST \citep{Ivezic08b,VanderPlas15}.   

For reference, Table 1 summarizes the areal coverage, effective RR Lyrae depth (distance of $\gtrsim$50\% completeness, in the absence of Galactic extinction), and maximum RR Lyrae distance (when known) for each of these surveys.   To calculate RR Lyrae distances from published apparent magnitudes, we assume [Fe/H] = -2.0 and M$_{\rm V}$ = 0.23([Fe/H]) + 0.93  which gives M$_{\rm V}  \sim$ 0.5 \citep{Chaboyer99}.  We also assume that M$_{\rm V} \sim$ M$_{\rm r}$,  as in \citet{Sesar13a}.  In many cases, the effective survey depth and maximum RR Lyrae distance are quite different.  This is because the RR Lyrae completeness and purity drop precipitously roughly a magnitude brighter than most survey's photometric limits.  However, in many cases \citep[e.g.][]{Drake13a,Drake13b}, interesting Galactic structure work can be done at magnitudes fainter than the effective RR Lyrae depth.

\begin{deluxetable*}{lcccccl}[t]
\tabletypesize{\scriptsize}
%\rotate
\tablecaption{Sensitivity of Time Domain Surveys to RR Lyrae Stars}
\setlength{\tabcolsep}{0.05in} 
\tablehead{ 
\colhead{Survey} &
\colhead{Coverage} &
\colhead{effective\tablenotemark{a}} &
\colhead{d$_{eff}\tablenotemark{a}$} &
\colhead{d$_{max}\tablenotemark{b}$} &
\colhead{Public RRL} &
\colhead{Completeness to RRab} \\
\colhead{} &
\colhead{deg$^2$} &
\colhead{mag limit} &
\colhead{kpc} &
\colhead{kpc} &
\colhead{Catalog} &
\colhead{and Comments}
}
\startdata
LINEAR & 10,000 & r$\sim$17.5 & $\sim$25 &  32 &  Y & $\sim$80\% for r$ <$17.2\\
SDSS-PanSTARRS-Catalina & 14,000 & g$\sim$17.8  & $\sim$28 & 37 &  Y & $\sim$50\% for V$<$17.8\\
CRTS CSS & $\sim$20,000 & V$\sim$18.0 & $\sim$30 & 80 & Y & $\sim$50\% at V = 18.0\\
CRTS SSS & 14,800 & V$\sim$18.0 & $\sim$30 & 60  & Y & $\sim$60\% at V = 18.0\\
CRTS MLS & $\sim$1,000 & V$\sim$20.0 & $\sim$75 & 125 & Y & $\sim$50\% at V = 20.0\\
SEKBO & 1675 & V$\sim$18.5 & $\sim$40 & 95 & Y & $\sim$60\% for V$<$18.5\\
La Silla QUEST & 840\tablenotemark{c} & r$\sim$19.0 & $\sim$50 & 95 & Y & $>$70\% for V $\le$19.\\
Stripe 82 & 270 & r$\sim$21.0 & $\sim$120 & 150 & Y & 99\% to 120 kpc \\

\hline

VVV & 562 & Ks$\sim$17 & $\sim$14 & -- & N & ongoing\\
PTF/iPTF/ZTF & 30,000\tablenotemark{d}  & r$\sim$20.0 & $\sim$80& -- & N & ongoing\\
PanSTARRS 1 & 30,000 & r$\sim$20.5 & $\sim$100 & -- & soon & data release in preparation \\
LSST Wide-Fast-Deep\tablenotemark{e} & $\sim$18,000 & r$\sim$24.5 & $\sim$600 & -- & TBD & future, 100\% complete at r$\sim$24.5  \\
& & & & & & images and photometry will be public                            
\enddata
\tablecomments{References: LINEAR - \citet{Sesar11,palaversa13}; PanSTARRS 1 - SDSS - \citet{Abbas14}; CRTS CSS/MLS - \citet{Drake09a}, \citet{Drake13a}, and \citet{Drake13b}; CRTS SSS - \citet{torrealba15a}; SEKBO - \citet{Keller08a}; LSQ - \citet{Zinn14a}; Stripe 82 - \citet{Sesar10a}; VVV - \citet{Saito12,gran15a}; PTF/iPTF/ZTF - \citet{Law09a}, \citet{Bellm14}, J. Surace (private communication, 2015); PanSTARRS 1 - B. Sesar (private communication 2015), Hernitschek et al. (in preparation); LSST- \citet{Ivezic08b}, \citet{VanderPlas15}}
\tablenotetext{a}{An effective magnitude and distance limit for RR Lyrae detection, based on the magnitude at which the RR Lyrae sample is reasonably complete, assuming no extinction and M$_{\rm V} = 0.5$.  ``Reasonably" ranges from $\sim$50\% to $\sim$90\% complete, depending on the information provided in the cited paper(s).}
\tablenotetext{b}{The maximum distance of a cataloged RR Lyrae, for surveys with publicly available catalogs.}
\tablenotetext{c}{840 deg$^2$ is the ``Region I" area discussed in \citet{Zinn14a}.  The LSQ will eventually survey 15,000 deg$^2$, http://hep.yale.edu/lasillaquest}
\tablenotetext{d}{The upper limit of the survey area to be covered by iPTF (J. Surace, private communication 2015) and/or ZTF.}
\tablenotetext{e}{We specify ``Wide-Fast-Deep" here to distinguish the part of the LSST survey that will be observed with a universal cadence, as described in \citet{Ivezic08b}, from other survey extensions such as Deep Drilling Fields and the Northern Ecliptic Spur. This is an underestimate of the area over which LSST data will be able to recover RR Lyrae.}
\end{deluxetable*}

\subsection{Fraction of Predicted Milky Way Sub-Halos Reached With RR Lyrae}\label{sec_elvis}

The time domain surveys that will be most relevant to the discovery of dwarf galaxies are those that are both wide-field and deep.  To identify the time domain surveys with the greatest potential to yield new Milky Way dwarf galaxy discoveries, we quantify the fraction of predicted dark matter sub-halos that will be contained within RR Lyrae survey volumes as a function of areal coverage and depth.  

We use the public\footnote{\url{http://localgroup.ps.uci.edu/elvis/}} ELVIS simulation suite \citep{GarrisonKimmel14a} to calculate the fraction of the dark matter sub-halos within 400 kpc that are enclosed within a given survey volume around the Milky Way, as a function of area and effective distance.  ELVIS includes a set of 12 high-resolution, dark matter only simulations of Local Group-like environments simulated in a cosmological context.  We made this prediction based on the average radial distribution of sub-halos around each of the Milky Way and M31 analogs in these 12 Local Group-like simulations.  We included sub-halos with v$_{\rm peak} >$ 12 km s$^{-1}$ and d$_{\rm host} <$ 400 kpc.  

\begin{figure}[htb!]
\plotone{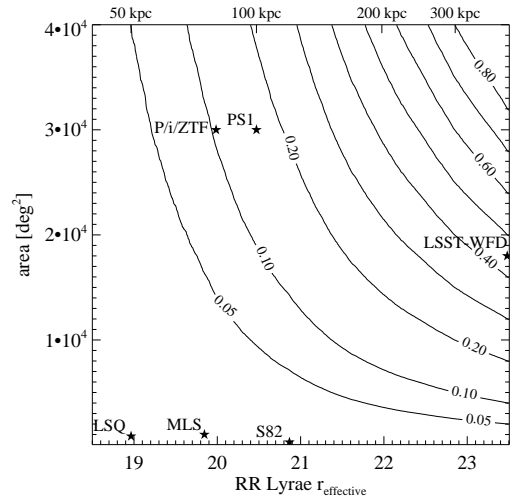}
\caption{The fraction of Milky Way sub-halos reached by RR Lyrae in current and planned time domain surveys, as a function of areal coverage and effective RR Lyrae detection distance.  The contours show lines of sub-halo fraction contained within a given survey area and depth. These survey volumes are overestimates, because they do not account for Galactic extinction close to the plane. However, LSST is expected to be much deeper than needed to find RR Lyrae out to the Milky Way's virial radius so this figures undersells the dwarf discovery volume to be enabled by LSST (at all but very low Galactic latitudes).}
\label{elvis}
\end{figure}

Figure~\ref{elvis} shows this prediction, with contours outlining the surfaces within which 5 -- 80\% of the predicted sub-halos are enclosed.  The effective volumes of the surveys listed in Table 1 are also overplotted. As in \S2.1, we assume M$_{\rm V} \sim$ M$_{\rm r} \sim$ 0.5 for RR Lyrae and no Galactic extinction to go from effective $V$ and $r$-depth to effective distance.  The estimated survey volumes in this section are therefore overestimates, because the effective RR Lyrae survey distances will be diminished in the Galactic plane.  However, these estimated fractions of sub-halos to be included within each survey volume are sufficient to guide our exploration of this approach to dwarf finding. Of pre-LSST surveys, both PTF/iPTF/ZTF and PanSTARRS 1 have the grasp to map $\sim$10-15\% of sub-halos with RR Lyrae stars.  

Only LSST will be sensitive to RR Lyrae occupying a significant fraction of Milky Way sub-halos.  LSST's universal wide-fast-deep (WFD) cadence will be sensitive to RR Lyrae well beyond the Milky Way's virial radius \citep[]{VanderPlas15, Oluseyi12a}, including $\sim$45\% of predicted sub-halos. Because LSST is expected to be much deeper than needed to reach the Milky Way's virial radius, this 45\% estimate will not be significantly diminished by extinction.  This is also a lower limit on the dwarf discovery space to be opened up by LSST, because it doesn't include (i) the volume between 400 kpc $<$ d $<$ 600 kpc that will also be mapped by RR Lyrae with the WFD cadence, (ii) thousands of additional square degrees to be observed with a non-universal cadence (e.g. the Northern Ecliptic Spur, Galactic Plane, and selected Deep Drilling Fields), and (iii) the fact that LSST's Deep Drilling Fields could reveal RR Lyrae to d $>$ 1.5 Mpc.

This Figure does not consider any special model for how dwarf galaxies are predicted to occupy sub-halos.  The fraction of Milky Way dwarf galaxies expected within these RR Lyrae survey volumes may, therefore, be somewhat different than the values shown in Figure~\ref{elvis}, even if we do reside in an LCDM universe.  \citet{hargis14a} used the ELVIS simulation suite to show that $\gtrsim$ 95\% of the Milky Way's dwarf galaxies are expected to reside at d $>$ 50 kpc, for several reasonable models used to populate dark matter sub-halos with dwarf galaxies.  However, the fraction of MW dwarf galaxies expected to reside at d $>$ 100 kpc ranges from 50\% - 80\%, depending on model.

%the distance beyond which $\gtrsim$ 50\% of the Milky Way's dwarfs are expected to reside varies with the model used to populate dark matter sub-halos with dwarf galaxies, and ranges from $\sim$100 - 170 kpc.  

This Figure also doesn't consider the detectability of any of these sub-halos (dwarf galaxies) as distinct groups of RR Lyrae stars.  In Section~\ref{sec_contam}, we will show that detecting dwarf galaxies as groups of RR Lyrae stars that are distinct from the field becomes easier at larger distances, where few field RR Lyrae are expected.  Therefore, deeper surveys for RR Lyrae will generally be more fruitful for dwarf galaxy discovery than wider surveys.

\section{RR Lyrae as a Tool for Dwarf Galaxy Discovery}\label{sec_sim}

There are two things that affect a dwarf galaxy's detectability as a group of RR Lyrae:  (i) whether a given dwarf galaxy has enough RR Lyrae stars with a compact enough spatial distribution to be identified as a group, and (ii) whether the number of interloper field RR Lyrae groups is low enough such that true dwarf galaxy groups are distinct.  To determine (i), we assume a population of RRab stars similar to those in known Milky Way dwarf galaxies (Section~\ref{sec_dwarfRRL}) and simulate the spatial distribution of RRab stars in dwarf galaxies with -10 $<$ M$_{\rm V}$ $< $-2.5 and 10 pc $<$ r$_{\rm half} <$ 1000 pc (Section~\ref{sec_simulate}).    To determine (ii), we apply a 2D friends-of-friends (FOF) group-finding algorithm (Section~\ref{sec_FOF}) to simulated RR Lyrae catalogs to study the expected demographics of interloper ``groups" of field RR Lyrae stars (Section~\ref{sec_contam}).   Informed by our investigation of the field RR Lyrae population, we apply an FOF algorithm with a linking length of 500 pc to the simulated dwarf galaxies to measure their detectability as a function of linking length, luminosity, and half-light size (Section~\ref{sec_detect}). 

%In Section~\ref{sec_detect}, we will synthesize these results to discuss the practical detectability of dwarf galaxies using RR Lyrae.

\subsection{Known RR Lyrae Populations of MW Dwarf Galaxies}\label{sec_dwarfRRL}

The Milky Way's dwarf spheroidal and ultra-faint dwarf satellites span four orders of magnitude in luminosity, including dwarfs with only several hundred Solar luminosities.    Despite the low luminosities of many Milky Way dwarfs, at least one RRab star has been found in every dwarf for which there is published time-series observations.  For example,  \citet{Simon11} and \citet{Boettcher13} discovered one RR Lyrae in each of Segue 1 and Segue 2 respectively, which are the two faintest MW dwarf galaxies currently known \citep{McConnachie12}.  On the other end of the MW dwarf spectrum, more than 100 RR Lyrae have been discovered in each of Draco, Leo II and Sculptor \citep{Bonanos04, Siegel00, Kaluzny95}.

\begin{figure}
\plotone{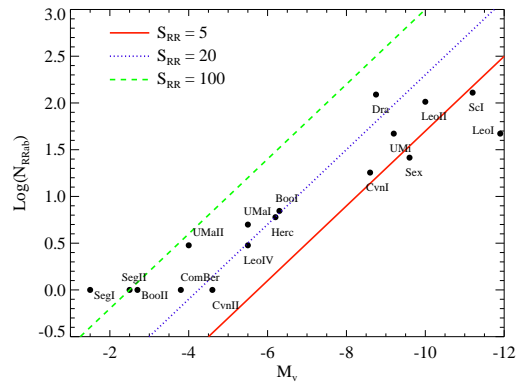}
\caption{The observed number of RRab stars in Milky Way dwarf galaxies as a function of V-band absolute magnitude. Three lines of constant specific frequency, $S_{\rm RR} = N_{\rm RR}*10^{0.4(7.5+ M_{\rm V})}$, are overplotted.  The specific frequencies of the most luminous dwarfs on this figure are lower limits, because of incompleteness.  Specific frequencies of 20 - 100 are the most representative of dwarf galaxies in the magnitude range we are most interested to find.}
\label{spec_freq}
\end{figure}

Table~\ref{dwarf_pops} summarizes the known number of RRab and RRc stars in 17 Milky Way dwarf galaxies and Figure~\ref{spec_freq} shows the observed number of RRab stars as a function of each Milky Way dwarf satellite's absolute V-band magnitude.  We focus Figure~\ref{spec_freq} on only RRab stars for three reasons: RRab are the most numerous type of RR Lyrae star,  RRab stars are easier to identify than RRc stars because of their distinctly shaped light curves and larger amplitudes, and RRab stars have well calibrated [Fe/H]-luminosity relationships.

\begin{deluxetable*}{lccccc}[t]
\tabletypesize{\scriptsize}
\tablecaption{Observed Milky Way Dwarf RR Lyrae Populations}
\setlength{\tabcolsep}{0.1in} 
\tablehead{ 
\colhead{MW Dwarf} &
\colhead{N$_{RRab}$} &
\colhead{N$_{RRc}$} &
\colhead{RR$_{c}$/RR$_{ab+c}$} &
\colhead{Reference} &
}

\startdata
Fornax & 396 & 119 & 0.23 & \citet{Bersier02}\\
Sculptor & 134 & 88 & 0.40 & \citet{Kaluzny95}\\
Draco & 214 & 30 & 0.12 & \citet{Kinemuchi08}\\
Leo II & 106 & 33 & 0.24 & \citet{Siegel00}\\
Ursa Minor & 47 & 35 & 0.43 & \citet{Nemec88}\\
Leo I & 47 & 7 & 0.13 & \citet{Held00}\\
Sextans & 26 & 10 & 0.27 & \citet{Mateo95}\\
Canes Venatici I & 18 & 5 & 0.22 & \citet{Kuehn08}\\
Bootes I & 7 & 8 & 0.53 & \citet{Siegel06}\\
Hercules & 6 & 3 & 0.33 & \citet{Musella12a}\\
Ursa Major I & 5 & 2 & 0.29 & \citet{Garofalo13}\\
Leo IV & 3 & 0 & 0.0 & \citet{Moretti09}\\
Coma Berenices & 1 & 1 & 0.5 & \citet{Musella09}\\
Canes Venatici II & 1 & 1 & 0.5 & \citet{Greco08}\\
Ursa Major II & 1 & 0 & 0.0 & \citet{DallOra12}\\
Segue 2 & 1 & 0 & 0.0 & \citet{Boettcher13}\\
Bootes II & 1 & 0 & 0.0 & \citet{Sesar14}\\
Segue 1 & 1 & 0 & 0.0 & \citet{Simon11}
\enddata
%\tablecomments{Observed RR Lyrae populations in known Milky Way Dwarfs}
\label{dwarf_pops}
\end{deluxetable*}

We overplot several values of specific frequency, using the parameterization of \citet{Mackey03}:

\begin{equation}
\rm S_{\rm RR} = N_{\rm RR}*10^{0.4(7.5+ M_{\rm V})}
\end{equation}

\noindent where N$_{\rm RR}$ is the number of RR Lyrae in the dwarf galaxy and specific frequency, S$_{\rm RR}$, is the number of RR Lyrae stars observed per 10,000 L$_{\odot
}$.  This Figure highlights a factor of 20 difference in observed specific frequency between the least luminous satellites (S$_{\rm RR} \sim$ 100) and the more luminous satellites (S$_{\rm RR} \sim$ 5).  Incompleteness in the RR Lyrae census in the brighter dwarf galaxies likely accounts for some of this difference.  Differences in stellar populations  may account for the rest of the difference.  For example, there is an observed stellar mass-metallicity relation and an observed stellar mass-SFH relation \citep[lower mass dwarfs seem to stop forming stars earlier, ][]{Kirby13a, Brown14a,Weisz15a}.

We adopt an S$_{\rm RR}$ of 20 - 100 to be representative of the low luminosity dwarf galaxies that RR Lyrae may help discover.  This S$_{\rm RR}$ gives a lower limit to the number of RR Lyrae that would be cataloged by studies that are also sensitive to RRc stars.    Known RRc fractions in the Milky Way's dwarfs range from $\sim$0.2 - 0.5.  The total number of RR Lyrae hosted by a dwarf galaxy is therefore 30\% - 100\% larger than expected for RRab stars alone.

\subsection{Simulating the RR Lyrae Populations of Dwarfs}\label{sec_simulate}

We simulated the RR Lyrae spatial distributions of 13,200 dwarf galaxies with -10 $<$ M$_{\rm V}$ $<$ -2.5 and 10 pc $<$ half-light radius $<$1000 pc.  Our idealized simulations assume 100\% RR Lyrae completeness.  In Section~\ref{sec_FOF}, we describe our simplistic RR Lyrae group finding approach.  It is based on a physical, rather than an angular, linking length so a dwarf's distance doesn't affect how detectable it is as a group of RR Lyrae.  (Although more distant dwarfs will be easier to pick out from the field RR Lyrae population - see Section~\ref{sec_contam}). 

13,200 dwarfs were simulated for each of two assumed RR Lyrae specific frequencies, S$_{\rm RR}$ = 20 and 100.  The number of RR Lyrae assigned to each simulated dwarf was drawn from a Poisson distribution with an average given by S$_{\rm RR}$/10$^{(0.4(7.5+ M_{\rm V}))}$.   The simulated stars were spatially distributed using an exponential distribution.  These simulations only account for the RRab population of dwarf galaxies, so underestimate the total number of RR Lyrae stars.

\subsection{Friends-of-Friends Approach to RR Lyrae Group Finding}\label{sec_FOF}

We apply a simplistic 2D FOF algorithm (IDL's {\it spheregroup}\footnote{\url{http://spectro.princeton.edu/idlutils_doc.html}}) for RR Lyrae group finding to each simulated dwarf, as well as to simulated MW stellar halo models (Section~\ref{sec_contam}) and public RR Lyrae catalogs (Section~\ref{sec_ObsGroups}).  We apply the algorithm on stars in overlapping bins of one apparent magnitude.  When initially investigating this method for detecting dwarfs, we tested 2D linking lengths between 50 pc and 500 pc (the physical size range observed for MW dwarf galaxies) to investigate the trade-off between increased recoverability of low surface brightness dwarfs as RR Lyrae groups and increased contamination by ``groups" of field RR Lyrae stars when using larger linking lengths.  We ultimately found that even a 500 pc linking length results in an acceptable level of field RR Lyrae group contamination at outer halo distances (Section~\ref{sec_contam}), so we will focus on the 500 pc linking length results for the rest of this paper.  

%As elsewhere in the paper, we assume M$_{\rm V}$ $\sim$ M$_{\rm r}$ $\sim$ 0.5 mag to go from apparent magnitude to distance - as necessary to use linking lengths in physical (rather than angular) units.  

\subsection{The Predicted Population of Interloper RR Lyrae Groups}\label{sec_contam}

The extent to which spatial associations of unbound field RR Lyrae stars will contaminate RR Lyrae-based searches for bound MW dwarf galaxies will be a function of distance, linking length, and the number of RR Lyrae required to define a group.   We focus our investigation of interloper RR Lyrae groups at d $>$ 50 kpc, both because there are a large number of field RR Lyrae with d $<$ 50 kpc and because $\sim$95\% of the MW's dark matter sub-halos are predicted to have d $>$ 50 kpc (Figure~\ref{elvis}). Although we tried linking lengths of 50 pc - 500 pc in our analysis, we only present the number of interloper field groups identified with a 500 pc linking length because, as we show below, that linking length yields a small number of interloper field RR Lyrae groups.

There is currently no observed sample of RR Lyrae stars that is both wide-field and deep enough to thoroughly investigate the demographics of field RR Lyrae at d $>$ 50 kpc (although we use the Stripe 82 catalog as a sanity check below).  We therefore turn to simulations to describe the expected properties of RR Lyrae in the MW's stellar halo.  (See Section 5 for caveats related to possible thick disk contamination in the Galactic plane.)

\citet{Lowing15} applied the semi-analytic galaxy formation model, GALFORM \citep{Cole00a,Font11a}, to five dark matter only simulations of Milky Way analogs.  They used the Aquarius A - Aquarius E halos from the Aquarius simulation project \citep{Springel08a}.  These are all relatively isolated halos with M$_{\rm tot}$ $\sim$ 10$^{12}$ M$_{\odot}$.  \citet{Lowing15} combined the particle tagging approach of \citet{Cooper10a} with stellar population synthesis and phase space sampling \citep[ENBID, ][]{Sharma06} to generate full catalogs of halo stars in each simulation.  Like any simulation, this method has some limitations. For example, it only accounts for the accreted population of halo stars, it doesn't include the lowest luminosity dwarf galaxy accretions (which contribute little to the halo), and stars are initially given the phase space distribution of the dwarf galaxy dark matter particles onto which they are tagged.  However, \citet{Lowing15} demonstrated the efficacy of these stellar halo catalogs to provide faithful realizations of the majority of the accreted stellar halo at d $>$ 20 kpc.  Moreover, we checked that the simulated z = 0 dwarf galaxy satellites have specific frequencies of RR Lyrae analog stars (as selected below) that are similar to those in observed dwarf galaxies. For the Aquarius B and D simulations, the S$_{\rm RR}$ of well resolved dwarfs with 50 kpc $<$ d $<$ 300 kpc ranged from $\sim$10 to 100, with most between 20 - 70 (a bit higher than observed for the Milky Way's more luminous dwarfs, as described in Section 3.1).

\begin{figure}
%\epsscale{0.8}
\plotone{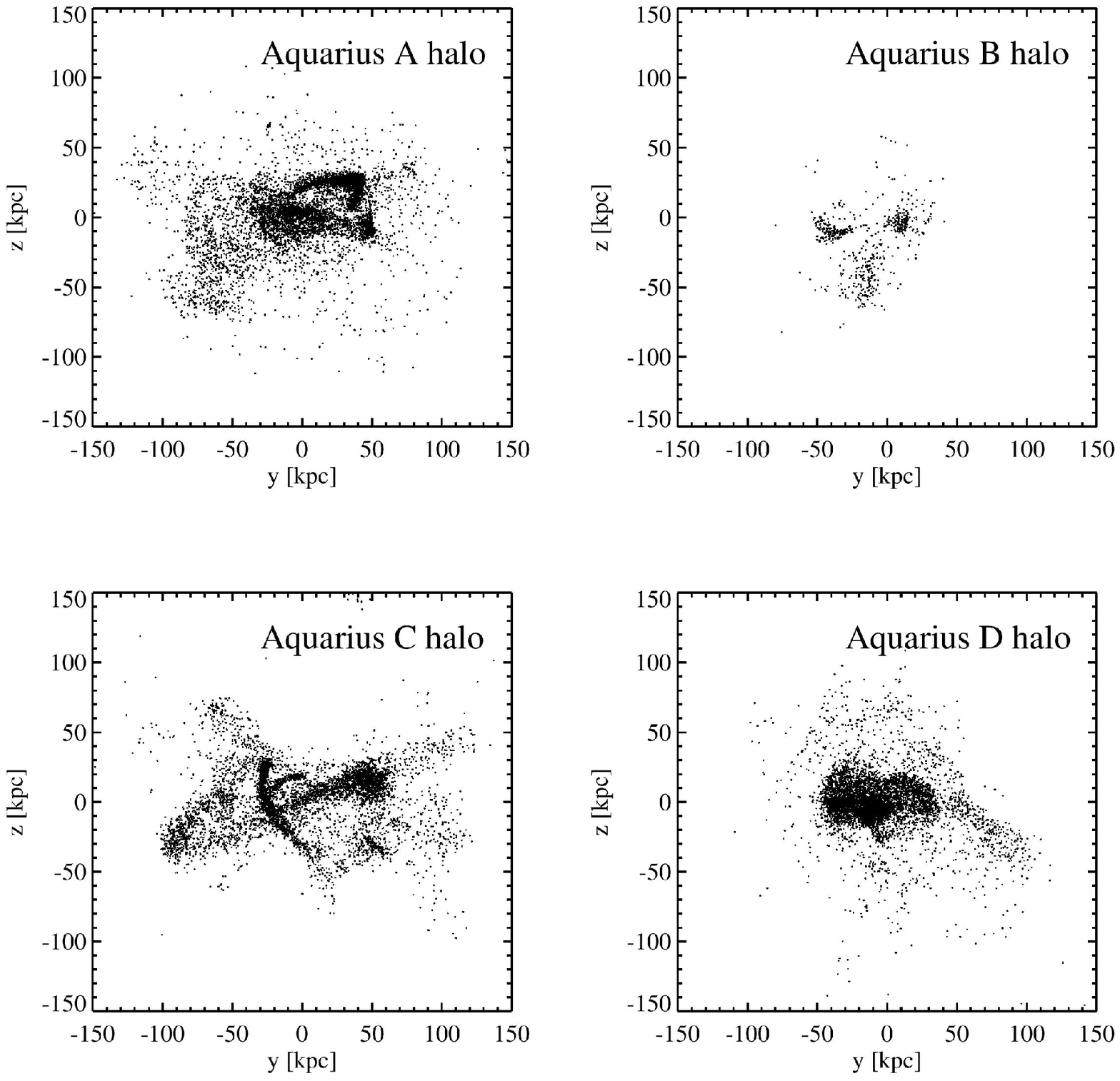}
\caption{The spatial distributions of RR Lyrae analog stars in semi-analytic stellar halo models applied to the Aquarius A - D simulations \citep{Springel08a,Lowing15}.  Stars were selected by applying criteria of (i) 6100 K $< T_{\rm eff} <$ 7400 K, (ii) 2.5 $<$ log(g) $<$ 3.0, (iii) unbound to any galaxy, and (iv) more than 50 kpc from the ``Sun" to the simulated stellar halo catalogs generated by \citet{Lowing15}.  The highly structured spatial distributions, and total number of RR Lyrae stars, are both in agreement with past predictions and current observations.}
\label{simulations}
\end{figure}

To select RR Lyrae analog stars from these catalogs\footnote{available at \url{http://galaxy-catalogue.dur.ac.uk:8080/StellarHalo/}}, we applied selection criteria of (i) 6100 K $< T_{\rm eff} <$ 7400 K, (ii) 2.5 $<$ log(g) $<$ 3.0, (iii) unbound to any galaxy, and (iv) more than 50 kpc from the ``Sun" \citep{Smith95}. The number of analog RR Lyrae stars satisfying these criteria ranges from $\sim$1,500 - 4,000 (Aquarius B and E) to $\sim$ 35,000 - 42,000 (Aquarius A, C, and D). These numbers are consistent with the number observed in Stripe 82, the only public RR Lyrae catalog that is complete to d $\sim$ 100 kpc.  There are 55 Stripe 82 RR Lyrae with 50 kpc $<$ d $<$ 100 kpc.  A crude area correction yields an expected $\sim$8,100 RRL in the MW with 50 kpc $<$ d $<$ 100 kpc, right in the middle of the predicted number in Aquarius A-E (622 to 39,738).

The spatial distributions of these simulated RR Lyrae field populations are also qualitatively similar to previous N-body+semi-analytic predictions of the stellar halos of MW-mass galaxies \citep[e.g.][]{johnston08a}.  Figure \ref{simulations} shows the spatial distributions of Aquarius A-D, in Cartesian coordinates, with the x-axis aligned with the major axis of the simulated dark matter halo.  (We exclude Aquarius E because it has a very small number of RR Lyrae with d $>$ 50 kpc and is qualitatively consistent with Aquarius B.)  These figures show that the predicted RRL distributions beyond 50 kpc are highly structured, with significant differences between the halos, and few RR Lyrae with d $>$ 100 kpc ($\lesssim$ one per 10 deg$^{2}$ on average, 965 -- 5,739 total).

We applied the FOF algorithm with a 500 pc linking length, as described in Section~\ref{sec_simulate}, to each of the five simulated stellar halos.   Depending on whether we implemented a group threshold of 2 or 3 RR Lyrae stars, N$_{\rm RR}$, each simulation contained dozens to a couple thousand field RR Lyrae ``groups" with d $>$ 50 kpc.  Although this sounds like a large number of field RR Lyrae groups, these groups are highly spatially clustered around a small number of rich stellar streams, with few isolated field RR Lyrae groups. The number of field RR Lyrae groups found in the simulations with 50 kpc $<$ d $<$ 100 kpc is consistent with the number found in the 280 deg$^2$ of Stripe 82.  Five groups of 2 stars are found in Stripe 82 and no groups were found with three or more stars.  (See Section~\ref{sec_ObsGroups} for more discussion). Assuming Stripe 82 is representative of the typical field halo RR Lyrae population around the Milly Way, the total number of groups expected  is $\sim$700. This is in between the small number found in Aquarius B and E ($\sim$200 - 300) and the number found in Aquarius A, C, and D ($\sim$2000 - 3000).
  
\begin{figure*}
\plottwo{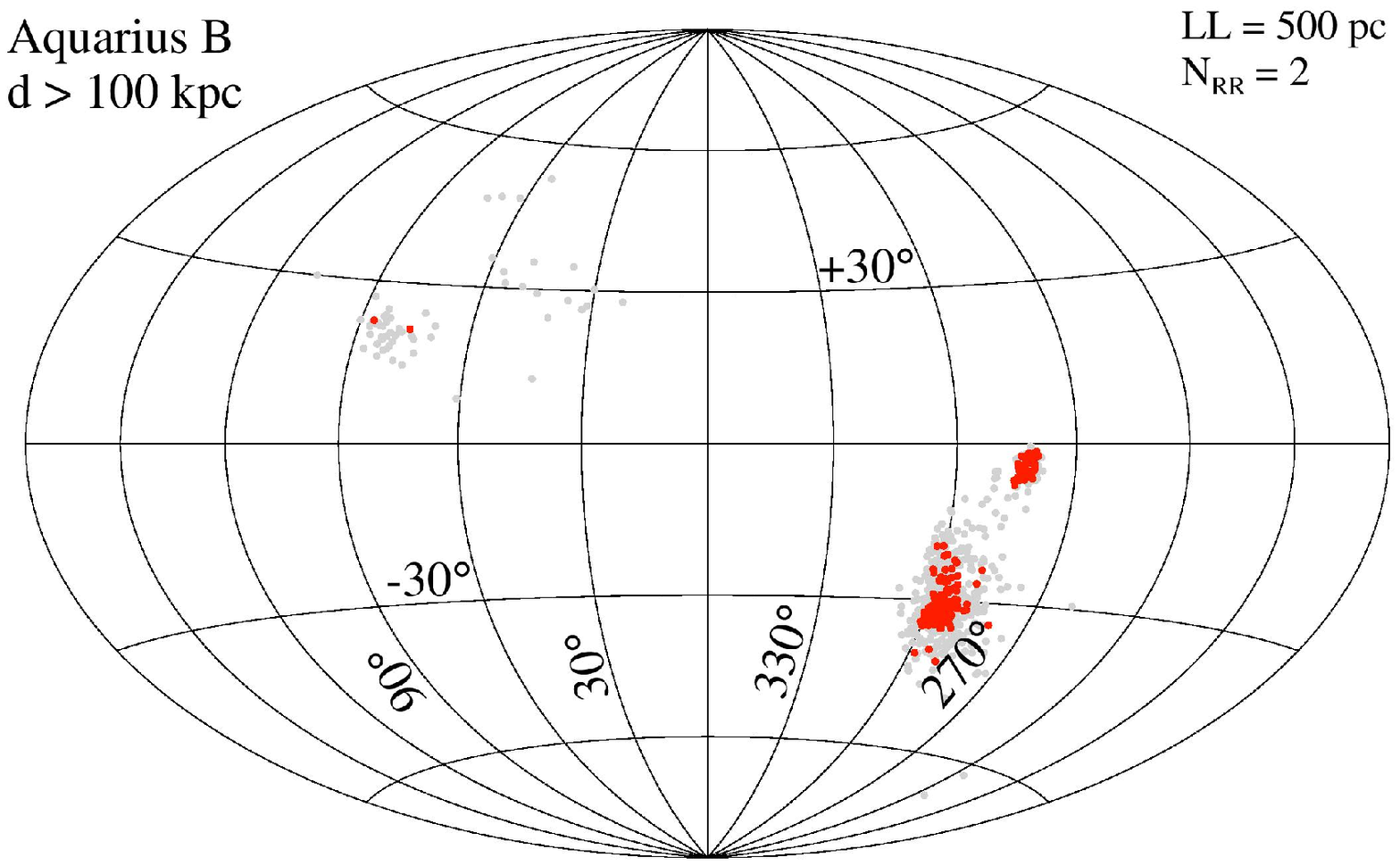}{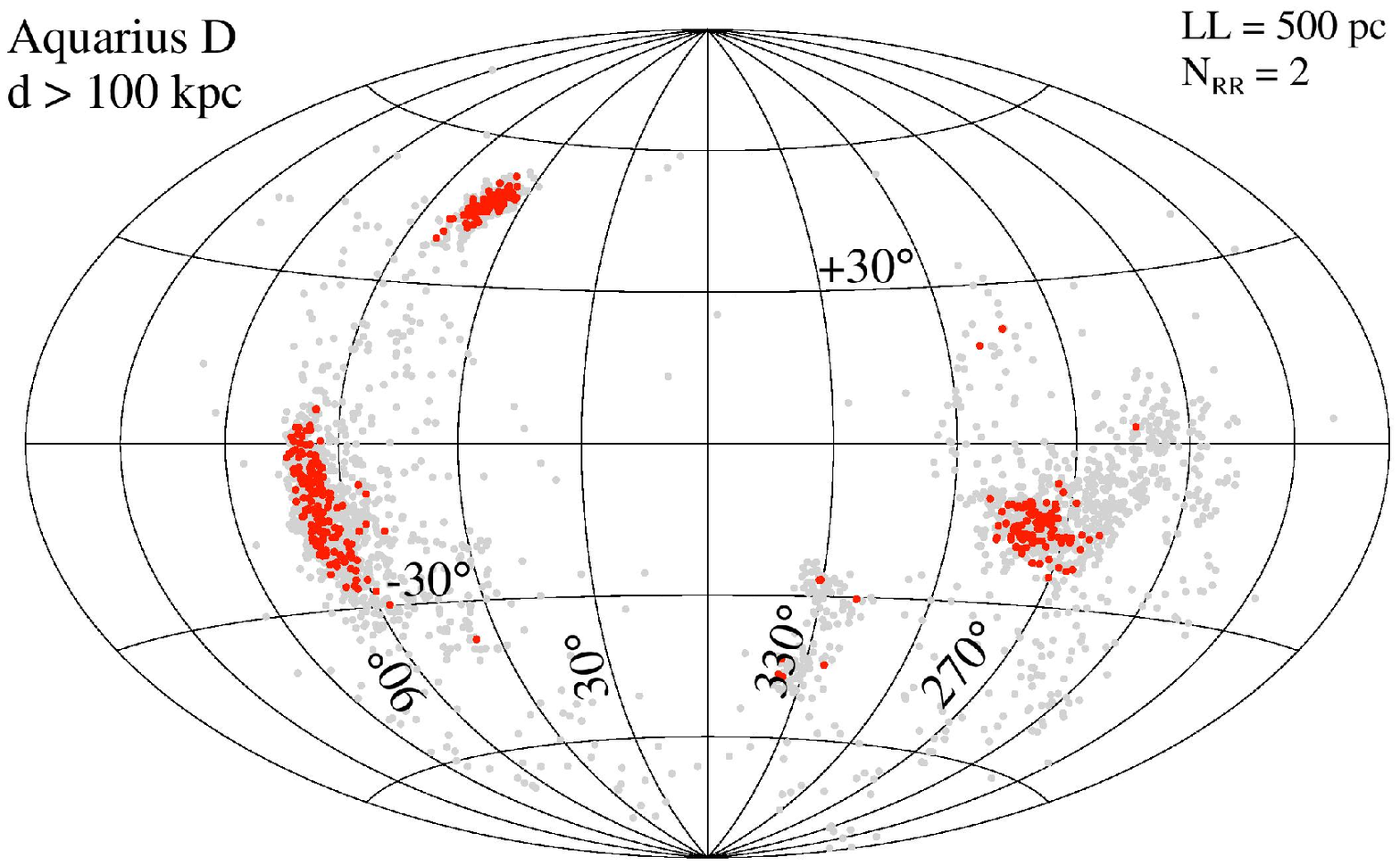}
\caption{The spatial distribution of RR Lyrae analog stars at d $>$ 100 kpc in semi-analytic stellar halo models applied to the Aquarius B and D simulations \citep{Springel08a,Lowing15}, shown in Equatorial coordinates.  Red symbols show friends-of-friends groups of two or more simulated field halo stars, identified with a linking length of 500 pc.  At these outer halo distances, very few isolated groups of field RR Lyrae stars are expected to contaminate searches for dwarf galaxies as groups of RR Lyrae stars.}
\label{contam1}
\end{figure*}

We first focus on the predicted number and spatial distribution of field RR Lyrae groups with d $>$ 100 kpc in the simulations, which includes the vast majority of the MW's virial volume and $\sim$80\%  of predicted dark matter sub-halos.  Figure~\ref{contam1} shows the distribution of such RR Lyrae analog stars in Aquarius B and Aquarius D in grey.  Groups identified with two or more RR Lyrae stars are overplotted in red.  At these large distances, there are fewer than 5 field RR Lyrae groups are isolated from obvious large-scale halo overdensities.  Dwarfs at d $>$ 100 kpc can therefore be identified with little contamination ($\lesssim$ 1 per 8000 deg$^2$ on average over large areas) as groups of 2 or more RR Lyrae linked within 500 pc, by ignoring RR Lyrae groups embedded near large-scale halo overdensities (which appear to occupy $<$1/6 of the sky).

\begin{figure*}
\plottwo{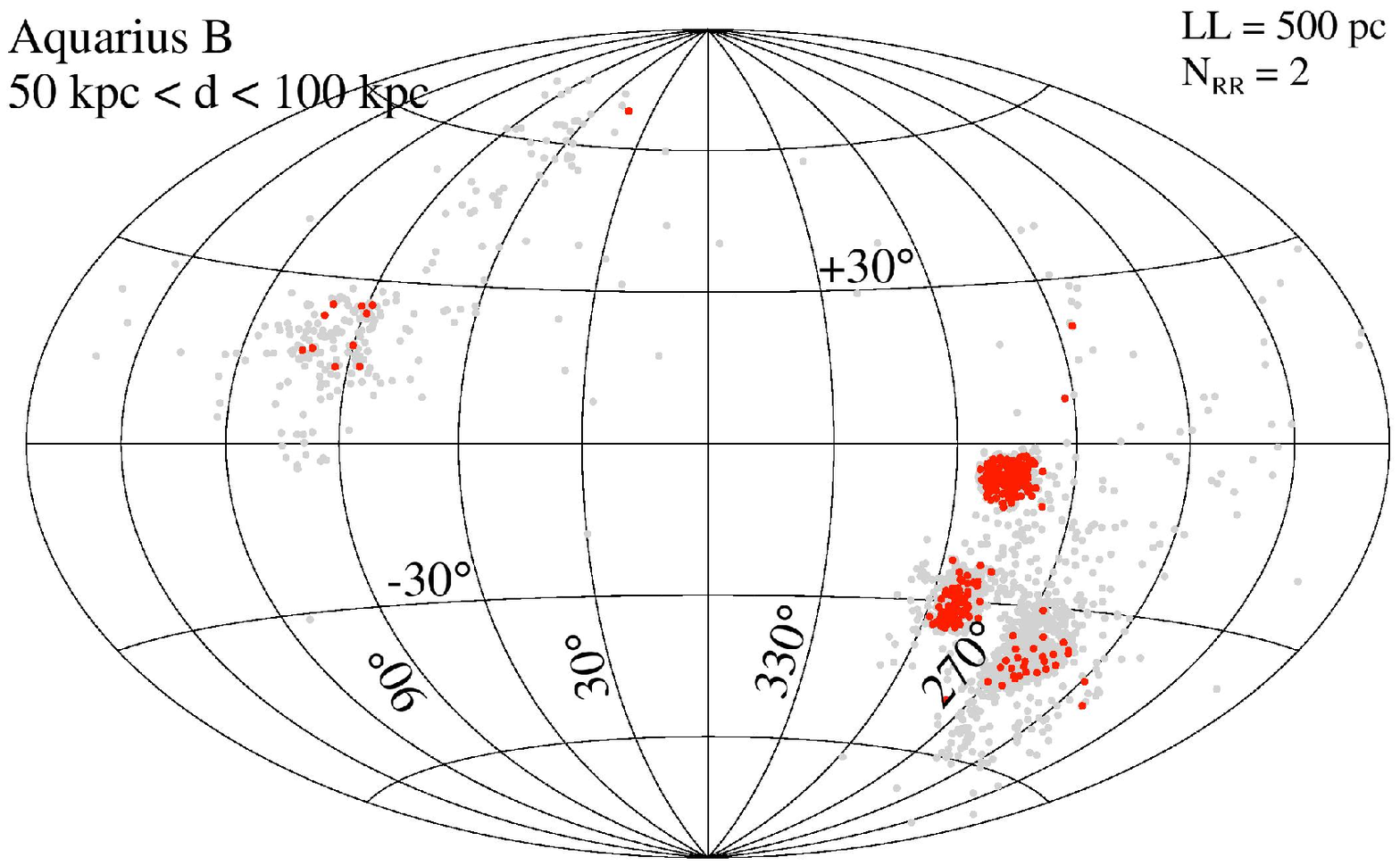}{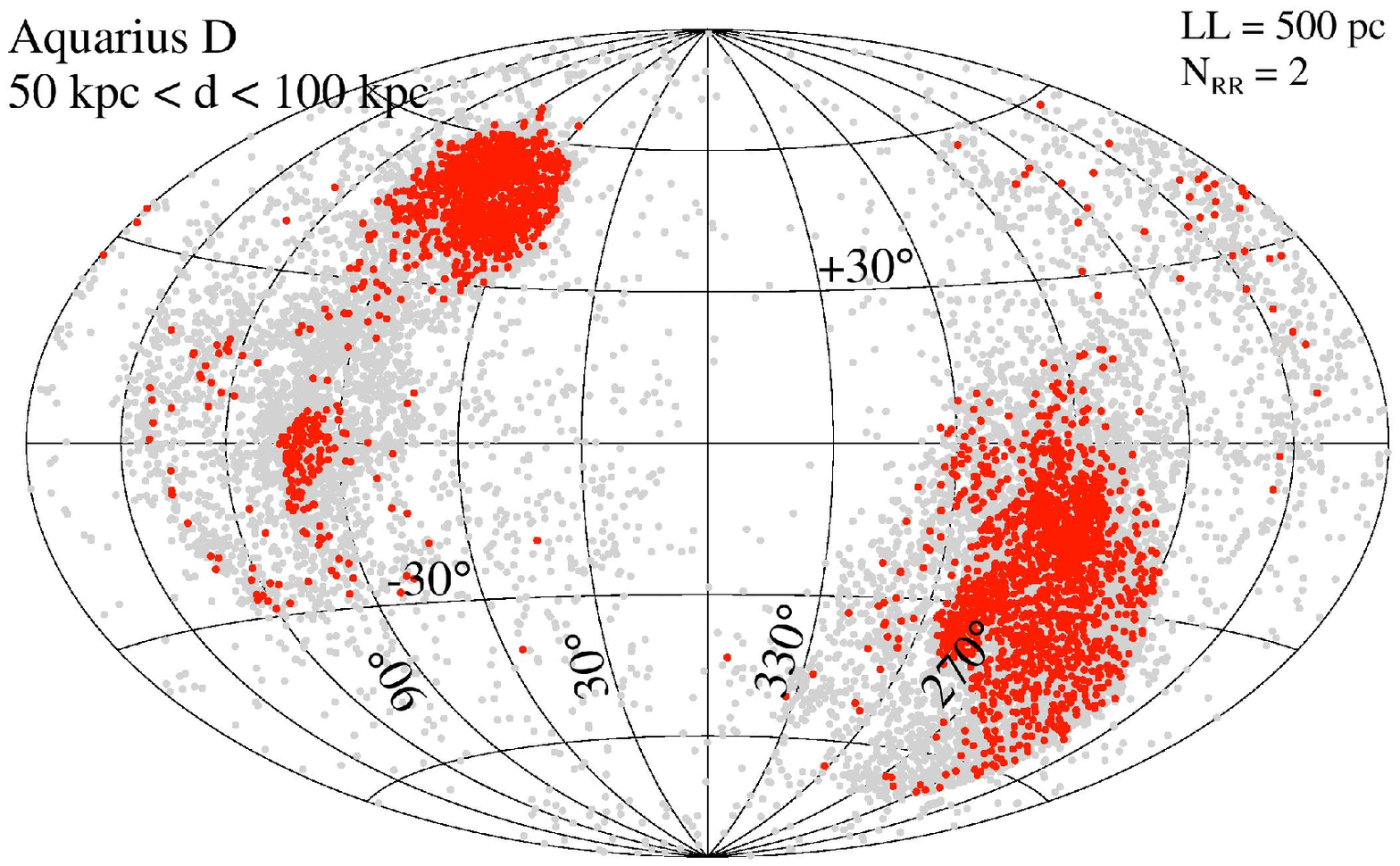}
\plottwo{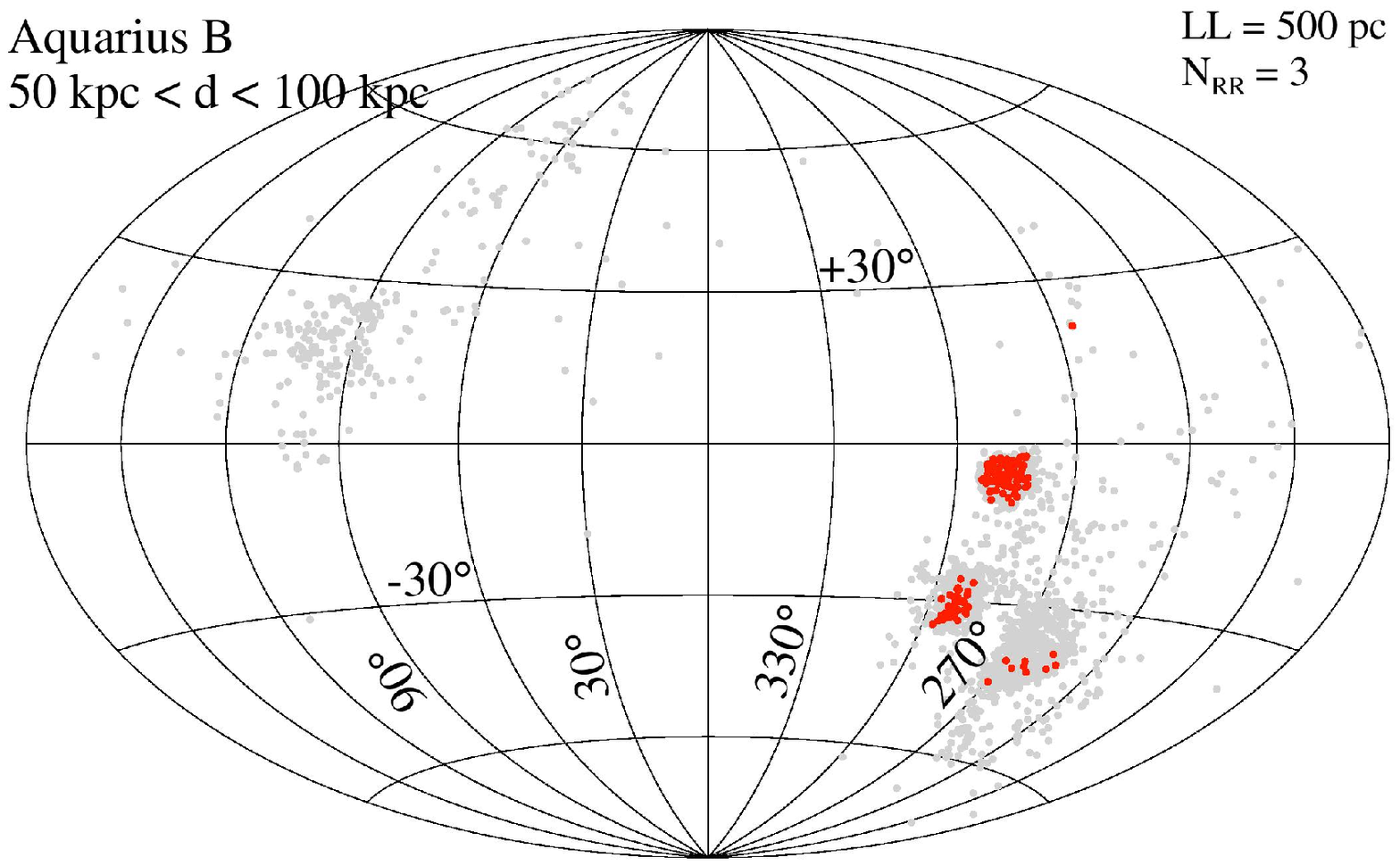}{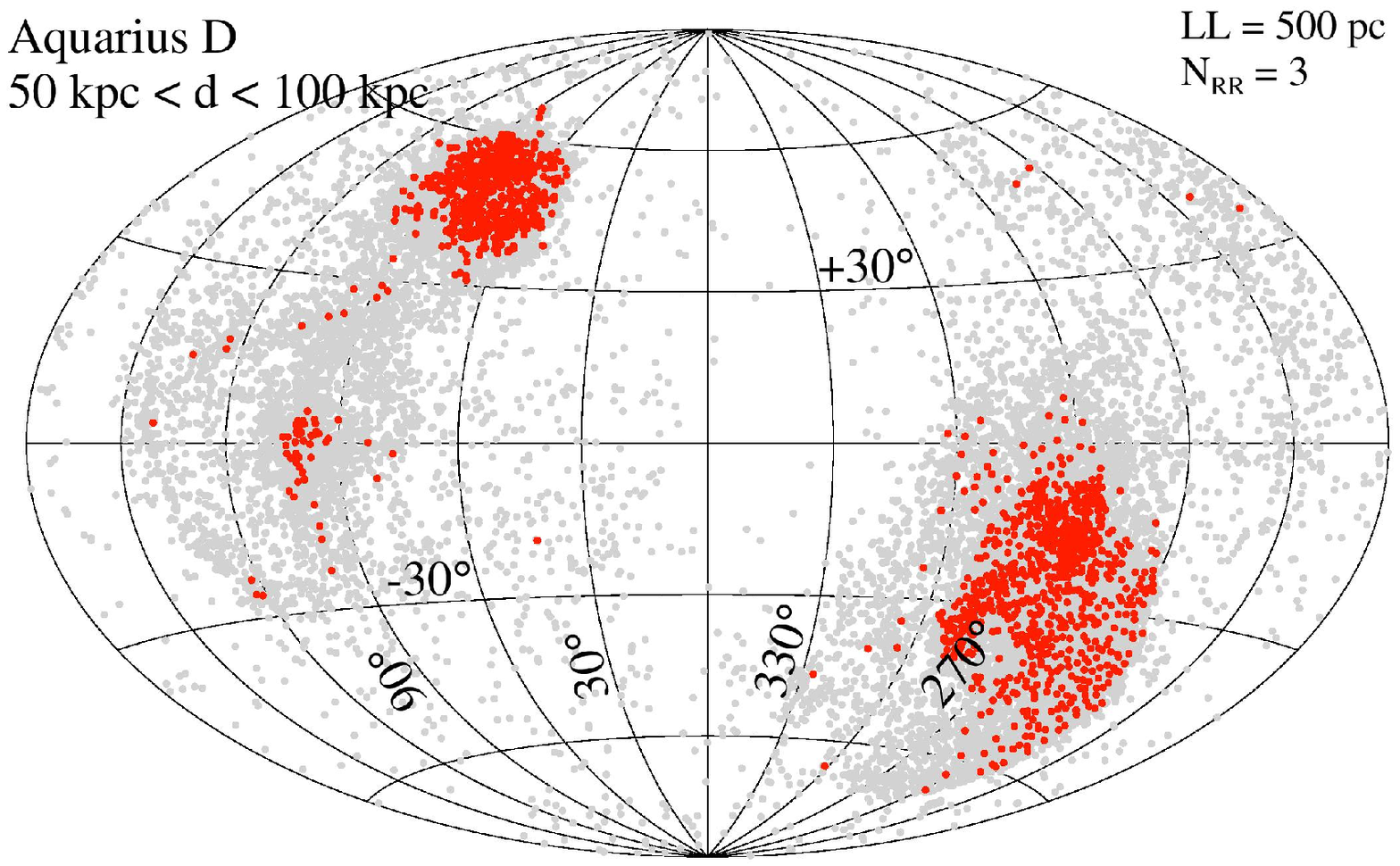}
\caption{{\it Top panels:} The spatial distribution of RR Lyrae analog stars at 50 kpc $<$ d $<$ 100 kpc in semi-analytic stellar halo models applied to the Aquarius B and D simulations \citep{Springel08a,Lowing15}, shown in Equatorial coordinates.  Red symbols show groups of two or more simulated field halo stars, identified with a linking length of 500 pc.  Although there are as many as a few dozen relatively isolated field groups or two or more RR Lyrae in this distance range, groups with only 2 stars will distinctly reveal new dwarf galaxies with 50 kpc $<$ d $<$ 100 kpc when quite isolated from other field RR Lyrae spatial clusterings.  {\it Bottom panels:} Same as top panels, but for groups of three or more RR Lyrae stars.  Fewer than 10 relatively isolated groups of field RR Lyrae are found with this higher threshold.  Dwarfs in this distance range could therefore be identified as groups of three or more RR Lyrae stars with $\lesssim$ 1 interloper group per 4000 deg$^2$, on average.}
\label{contam2}
\end{figure*}

We next focus on the predicted number and spatial distribution of field RR Lyrae groups with 50 kpc $<$ d $<$ 100 kpc, which includes a few percent of the MW's virial volume and $\sim$15\% of predicted dark matter sub-halos.  Figure~\ref{contam2} is the same as Figure~\ref{contam1} but for 50 kpc $<$ d $<$ 100 kpc and shows that many more field RR Lyrae stars exist with 50 kpc $<$ d $<$ 100 kpc, with large-scale halo overdensities occupying up to 1/2 of the sky in projection.  The top two panels of Figure~\ref{contam2} show the distribution of groups with two or more RR Lyrae and the bottom two panels show the distribution of groups with three or more RR Lyrae.  There are as many as a few dozen relatively isolated groups found with two or more RR Lyrae, but less than 10 relatively isolated groups of three or more RR Lyrae.   With fewer than 10 relatively isolated field RR Lyrae groups, dwarfs at 50 kpc $<$ d $<$ 100 kpc could be identified with little contamination ($\lesssim$ 1 per 4000 deg$^2$ on average over large areas) as groups of 3 or more RR Lyrae linked within 500 pc, by ignoring RR Lyrae groups embedded near large-scale halo overdensities (which occupy $<$1/2 of the sky).  Groups with only 2 stars can also reveal new dwarf galaxies with 50 kpc $<$ d $<$ 100 kpc when quite isolated from other field RR Lyrae clusterings.

%However, the 500 pc linking length combined with N$_{\rm RR}$ = 3 opens up significantly more discovery space than the 50 pc linking length combined with N$_{\rm RR}$ = 2.  The bottom two panels of Figure~\ref{contam} therefore show the results of an FOF search with linking length = 500 pc, N$_{\rm RR}$ = 3 on Aquarius B and D RR Lyrae analog stars with 50 kpc $<$ d $<$ 100 kpc. 

\subsection{The Detectability of Milky Way Dwarfs Using RR Lyrae}\label{sec_detect}

We focus our discussion of detectability on dwarfs that would be identified as groups of two or more RR Lyrae stars, with a 2D linking length of 500 pc.  In Section~\ref{sec_contam}, we demonstrated that we expect few isolated, field RR Lyrae groups of two of more RR Lyrae in the outer halo (d $>$ 100 kpc).  These distances include the vast majority of the MW's halo volume and predicted dwarf galaxies.  Groups of two or more RR Lyrae stars at nearer distances (50 kpc $<$ d $<$ 100 kpc) that are quite isolated from visible RR Lyrae structures will also distinctly identify dwarf galaxies.  In the discussion below, the luminosity detection limits would be $\sim$0.5 mags brighter with a threshold of three RR Lyrae, as might be desired at these nearer distances (50 kpc $<$ d $<$ 100 kpc) to reduce the number of expected field RR Lyrae groups that are spatially coincident with diffuse halo structures.  

\begin{figure*}
\epsscale{0.5}
\plotone{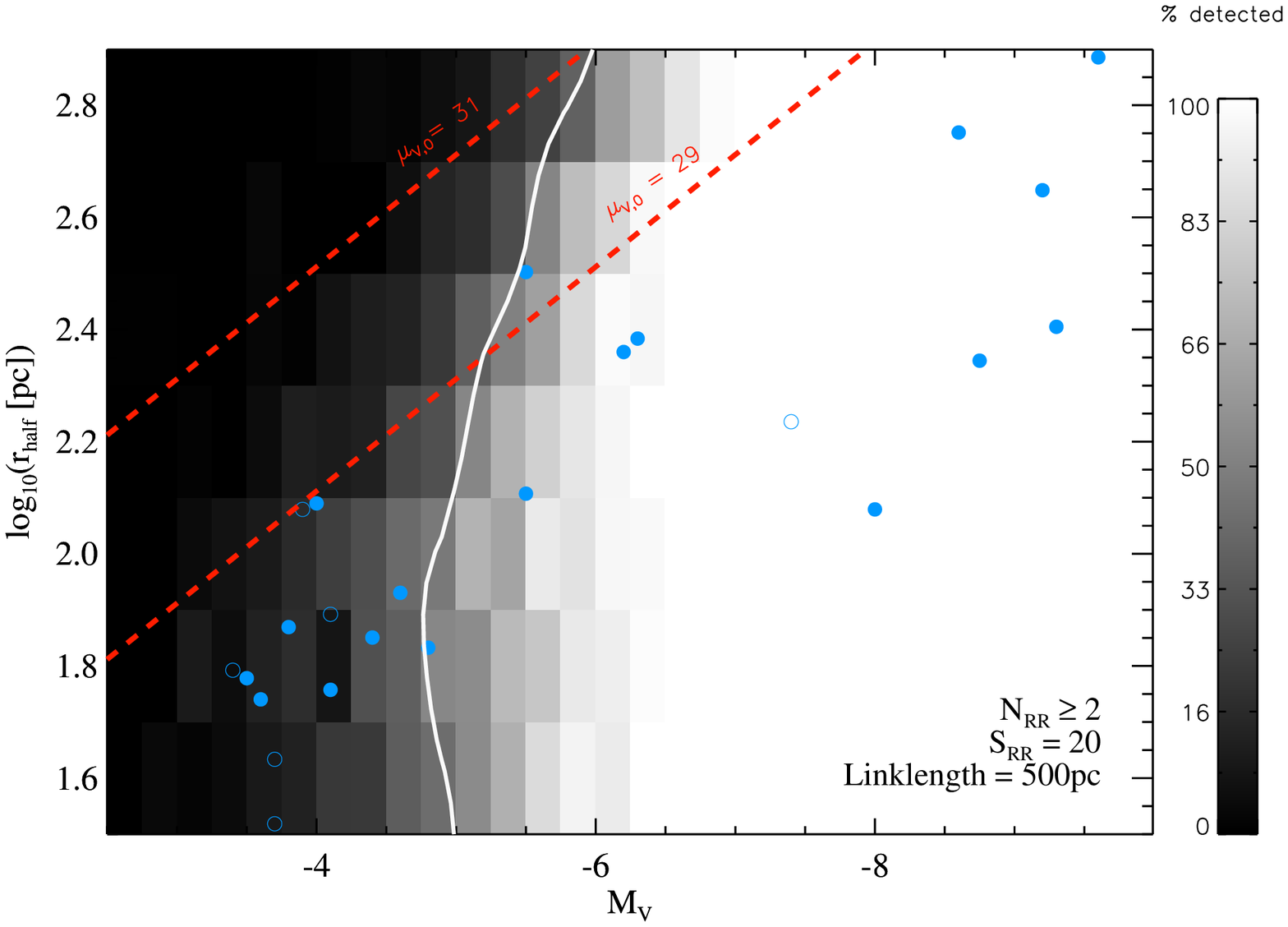}
\plotone{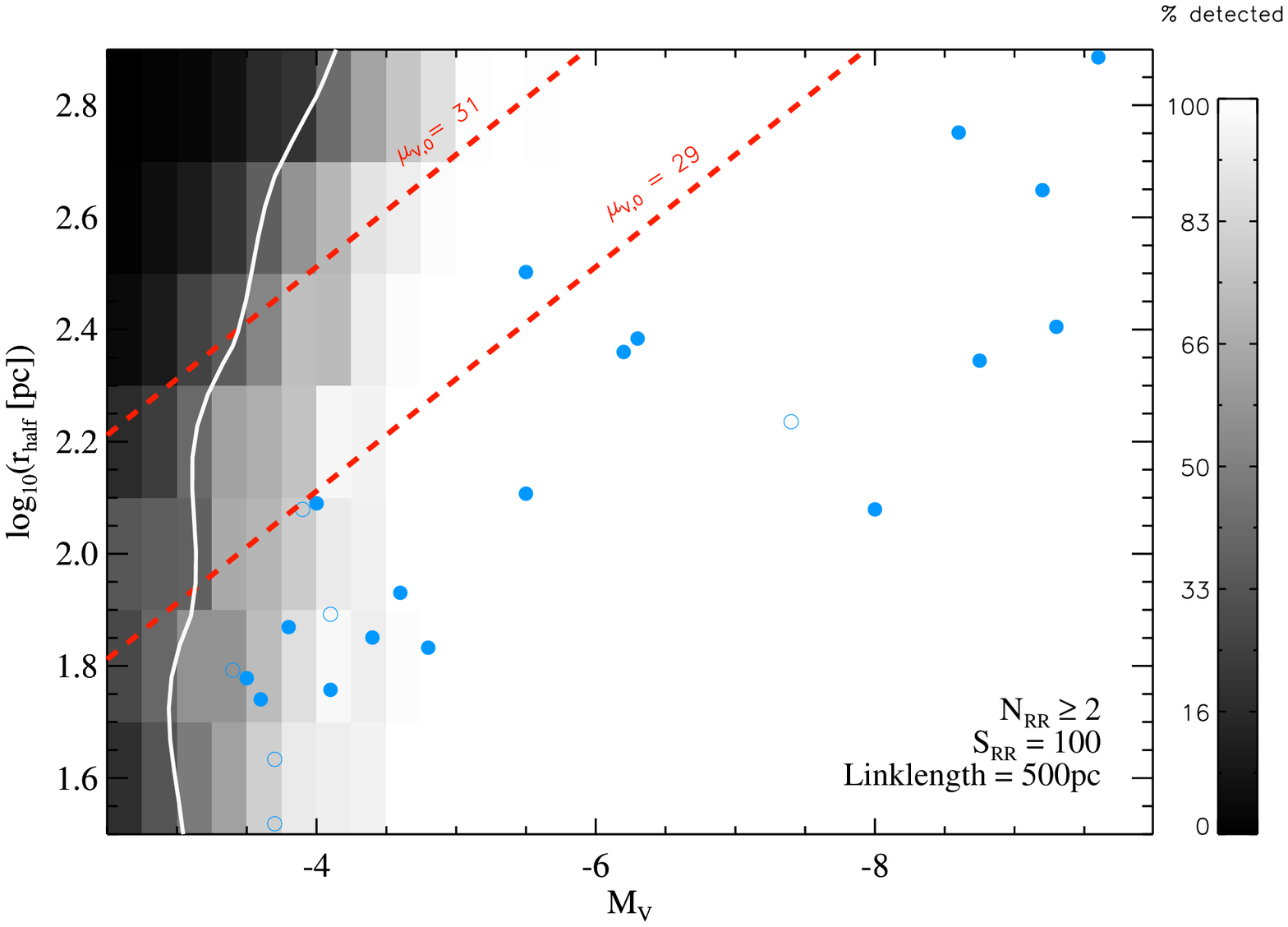}
\caption{The detectability of simulated dwarfs as groups of 2 or more RR Lyrae stars identified with a 500 pc linking length, as a function of size, M$_{\rm V}$, and RR Lyrae specific frequency, S$_{\rm RR}$.  A group threshold of 2 is most appropriate for d $>$ 100 kpc, or for 50 kpc $<$ d $<$ 100 kpc if some contamination in the sample of dwarf candidates is acceptable or if groups found near large overdensities are excluded (see Figure~\ref{contam2}).  Blue circles show known (filled points) and candidate (open points) MW dwarfs.  The white line traces the sizes and luminosities of dwarfs that are detected as RR Lyrae groups 50\% of the time.  The results in the left (right) panel assumes an RRab star specific frequency of 20 (100).  Dwarfs with relatively lower specific frequencies are likely those more luminous than M$_{\rm V}$ = -4.5 mag (see Section~\ref{sec_dwarfRRL}).  Identifying dwarfs as groups of 2 or more RR Lyrae stars will open up significant discovery space at low Galactic latitudes and low surface brightnesses.}
\label{detectability2a}
\end{figure*}

Figure~\ref{detectability2a} shows the detectability of simulated dwarfs as groups of two or more RR Lyrae stars, as a function of size, absolute magnitude, and RR Lyrae specific frequency, S$_{\rm RR}$.   The white line traces the sizes and luminosities of dwarfs that are identified as RR Lyrae groups 50\% of the time.  The left (right) panel shows the detectability of dwarfs with S$_{\rm RR}$ = 20 (100).  These specific frequencies are similar to those observed in dwarfs more (less) luminous than M$_{\rm V}$ $\sim$ -4.5 mag (see Section~\ref{sec_dwarfRRL}).  In fact, all dwarfs less luminous than M$_{\rm V} \sim$ -6 have RRab-only specific frequencies of at least 20.  The simulations informing the left panel therefore include the fewest RR Lyrae stars expected in ultra-faint dwarfs, yielding conservative detection limits at those luminosities. 

%For a group threshold of 2 RR Lyrae, specific frequencies of S$_{\rm RR}$ = 20 (100) should yield an absolute detectability threshold of M$_{\rm V}$ $\sim$\ -5.0 (-3.2) mag - the absolute magnitudes that produce an average of 2 RR Lyrae.    These limits are seen, as expected, in Figure~\ref{detectability2a}.  

For a specific frequency of 20, dwarfs similar to those known, with M$_{\rm V}$ brighter than $\sim$ -5 mag, will be identified as groups of 2 or more RR Lyrae stars.  Assuming S$_{\rm RR}$ = 20, low surface brightness dwarfs (at least as faint as $\mu_{\rm V,0}$  $\sim$31 mag arcsec$^{\rm -2}$) that are more luminous than M$_{\rm V}$ = -6 mag will also be detectable as groups of two or more RR Lyrae stars.  Even with our conservative estimate of the number of RR Lyrae to belong to these dwarfs, this demonstrates that groups of two or more RR Lyrae will distinctly reveal (i) dwarfs brighter than M$_{\rm V}$ = -5 mag that are similar to those known, but at low Galactic latitudes and (ii) slightly more luminous dwarfs at previously unexplored surface brightnesses.

%As discussed in Section~\ref{sec_dwarfRRL}, known dwarfs with -8 $<$ M$_{\rm V}$ $<$ -2 have RR Lyrae populations consistent with specific frequencies, S$_{RR}$, $\sim$ 20 - 100, with more luminous dwarfs displaying specific frequencies on the lower end of that range. Therefore, the left S$_{\rm RR}$ = 20 panels are most appropriate for dwarfs more luminous than M$_{\rm V}$ = -4.5 and the right S$_{\rm RR}$ = 100 panels are the most appropriate for dwarfs less luminous than M$_{\rm V}$ = -4.5.  

The detectability of dwarfs with luminosities fainter than M$_{\rm V}$ = -5 mag will depend sensitively on their RR Lyrae specific frequency.  Assuming a specific frequency of 100 for the least luminous dwarfs (e.g. M$_{\rm V}$ less luminous than -4.5 mag), the right panel demonstrates that dwarfs similar to those known and with luminosities as low as M$_{\rm V}$ = -3.2 mag will be detected as groups of two or more RR Lyrae stars.  Dwarfs will also be detected at these low luminosities, with surface brightnesses as faint as 31 mag arcsec$^{-2}$.   This demonstrates that groups of two or more RR Lyrae will also distinctly reveal dwarfs with -4.5 mag $<$ M$_{\rm V}$ $<$ -3.2 mag that are (i) at low Galactic latitudes and/or (ii) at previously unexplored surface brightnesses (depending on RR Lyrae specific frequency).

Although RR Lyrae will open up significant new dwarf discovery space, there are some dwarfs that RR Lyrae will not efficiently detect.  These include those less luminous than M$_{\rm V}$ = -3.2 mag unless they have an RR Lyrae specific frequency higher than yet observed, dwarfs fainter than M$_{\rm V}$ = -5 mag with specific frequencies less than 20, and dwarfs with -6 mag $<$ M$_{\rm V}$ $<$ -5 mag and lower surface brightness than 31 mag arcsec$^{\rm -2}$.

\section{Searching for RR Lyrae Groups in Public Survey Data}\label{sec_ObsGroups}

Although most published public RR Lyrae catalogs are severely incomplete for d $>$ 50 kpc (and all but Stripe 82 are severely incomplete for d $>$ 100 kpc), several of them include RR Lyrae with d $>$ 80 kpc (CRTS - the CSS and MLS, SEKBO, LSQ, and Stripe 82 - see Table 1 and references therein).  It is worth searching these catalogs for isolated groups of RR Lyrae, because very low surface brightness or low Galactic latitude dwarfs with the luminosity of (for example) Sextans or Draco, could be identified as an RR Lyrae group even with 10\% RR Lyrae completeness.  We therefore apply the FOF group finder with a 500 pc linking length, as described in Section~\ref{sec_FOF}, to these catalogs.  As before, we only consider RR Lyrae with d $>$ 50 kpc and, when possible, we included only cataloged RRab stars.  To consistently estimate distances to the RR Lyrae in each of these surveys, we assumed M$_{\rm V}$ $\sim$ M$_{\rm r}$ = 0.5 mag.  We also applied \citet{sfd98} corrections for dust extinction to the LSQ and CRTS magnitudes, and applied the correction for dust extinction as provided in the SEKBO catalog.  The Stripe 82 RR Lyrae catalog provided with an extinction correction applied. 

%We used the catalogs available for these surveys at \url{http://nesssi.cacr.caltech.edu/DataRelease/Varcat.html}; \url{http://vizier.cfa.harvard.edu/viz-bin/VizieR?-source=J/ApJ/678/851}; \url{http://iopscience.iop.org/0004-637X/781/1/22/fulltext/apj489349t1_mrt.txt} and \url{http://vizier.cfa.harvard.edu/viz-bin/VizieR?-source=J/ApJ/708/717}; \url{http://vizier.cfa.harvard.edu/viz-bin/VizieR-3?-source=J/MNRAS/424/2528/table1}, respectively.

\begin{figure}
\plotone{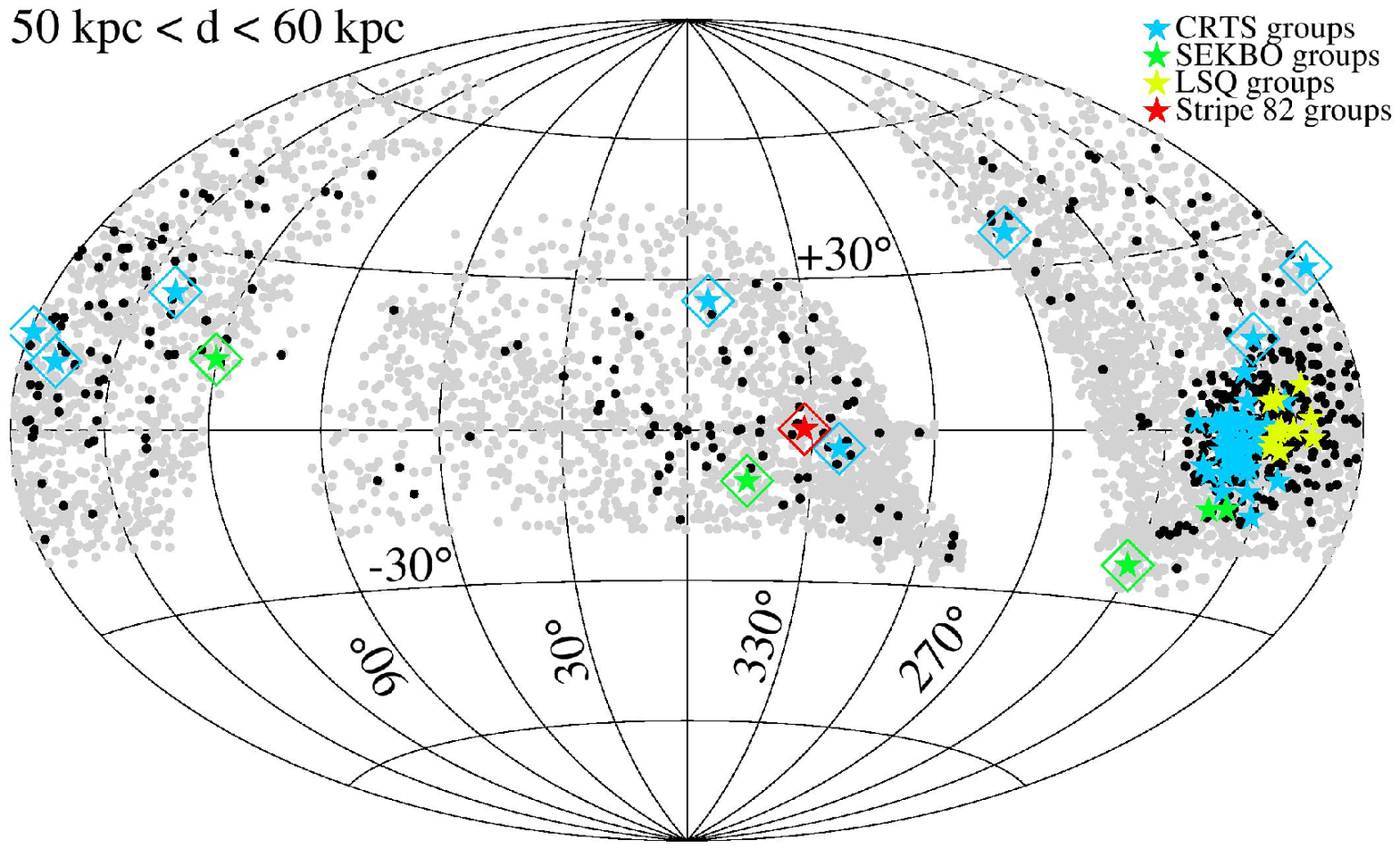}
\plotone{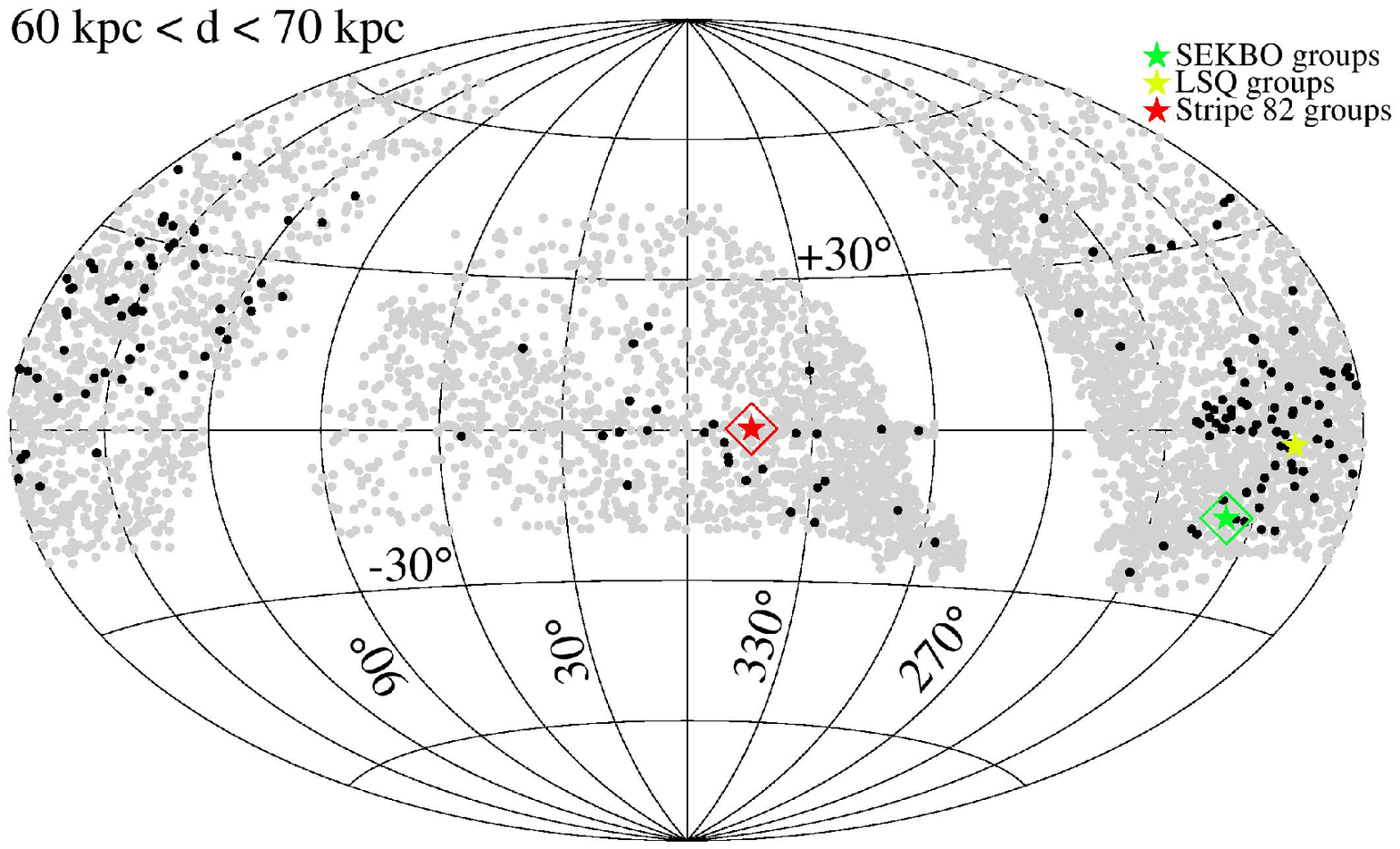}
\plotone{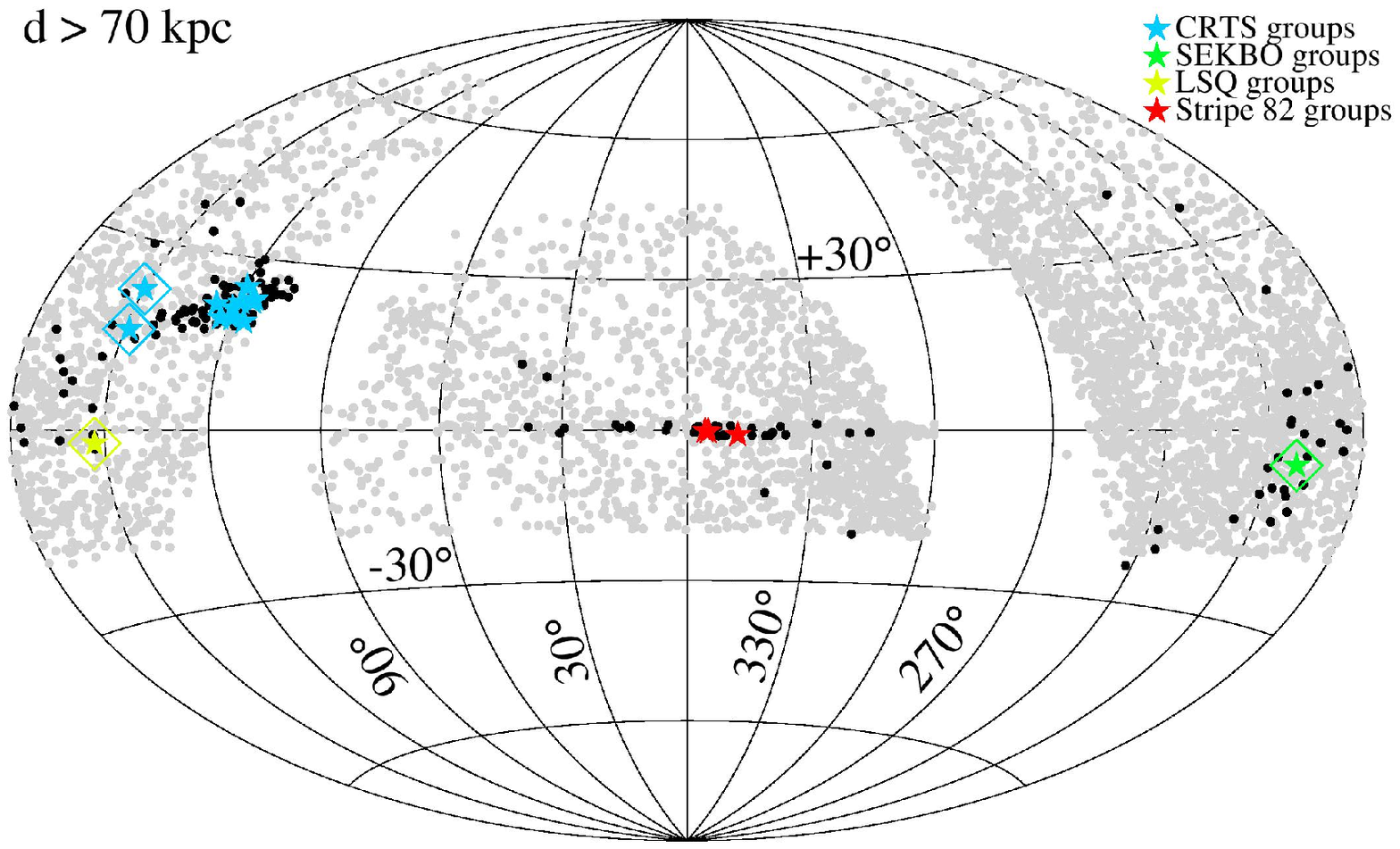}
\caption{Spatial distribution of RR Lyrae in  the CRTS, LSQ, SEKBO, and Stripe 82 public catalogs, in Equatorial coordinates.  The grey symbols in all three panels demonstrate the combined survey footprint, with the spatial distribution of all RR Lyrae in these surveys.  Each panel also shows RR Lyrae (black symbols) and groups or two or more RR Lyrae (colored symbols) in different distance slices.  Groups liberally classified as "relatively isolated" have boxes around them; 14 of of these likely belong to known halo structures, two to the Bo\"otes I and Sextans dwarf, and two are unknown.  {\it Top panel:} 50 kpc $<$ d $<$ 60 kpc; {\it Middle panel:} 60 kpc $<$ d $<$ 70 kpc; {\it Bottom panel:} d $>$ 70 kpc).  To consistently estimate distances to the RR Lyrae in all of these surveys, we assumed M$_{\rm V}$ $\sim$ M$_{\rm r}$ = 0.5 mag.  There are thus modest differences between the distances shown here and shown in the RR Lyrae survey papers.}
\label{aitoff_plots}
\end{figure}

Figure~\ref{aitoff_plots} shows the spatial distributions of RR Lyrae and RR Lyrae groups in the CRTS, LSQ, SEKBO, and Stripe 82 catalogs.  Grey symbols show the spatial distribution of all RR Lyrae in these catalogs, black symbols show the RR Lyrae contained within each of three distance bins (50 kpc $<$ d $<$ 60 kpc, 60 kpc $<$ d $<$ 70 kpc, and d $>$ 70 kpc), and colored symbols show groups of two or more RR Lyrae.  

%Motivated by the results of Section~\ref{sec_contam}, we selected RR Lyrae groups that appear relatively isolated from significant RR Lyrae overdensities.  We chose to be liberal in classifying groups as relatively isolated, to be as inclusive of potentially interesting structures as possible.  These are outlined in boxes and listed in Table~\ref{final_table}.   We highlight known halo structures (Section~\ref{sec_known_structures}) and discuss these relatively isolated groups (Section~\ref{sec_new_groups}) below.

\begin{deluxetable*}{cccccccc}
\tabletypesize{\scriptsize}
%\rotate
\tablecaption{Relatively isolated RR Lyrae groups in public catalogs}
\setlength{\tabcolsep}{0.1in}
\tablehead{ 
\colhead{RA} &
\colhead{Dec} &
\colhead{Avg M$_{\rm v}$} &
\colhead{Distance\tablenotemark{a}} &
\colhead{N$_{\rm RR}$} &
\colhead{Size (pc)\tablenotemark{b}} &
\colhead{Survey} &
\colhead{Classification}
}

\startdata

153.17955 &   -2.04230 &  20.2 &    84 &     2 & 328 &     LSQ & Sextans\\
 209.85728 &   14.40900 &  19.4 &    59 &     4 & 674 &    CRTS & Bootes I\\
\hline
120.40614 &   12.16494 &  19.3 &    57 &     2 & 150 &   SEKBO & likely Sgr\\
140.01188 &   22.67615 &  19.3 &    57 &     2 & 513 &   CRTS & likely Sgr\\
168.27063 &   10.05624 &  19.1 &    52 &     2 & 505 &   CRTS & likely Sgr\\ 
178.85546 &   13.87637 &  19.4 &    59 &     2 & 107 &  CRTS & likely Sgr\\
180.65865 &   23.47226 &  19.4 &    58 &     2 & 406 &   CRTS & possible Sgr?\\
218.35120 &  -14.26636 &  19.5 &    61 &     2 & 172 &    SEKBO & possible Sgr?\\
241.24207 &  -23.50549 &  19.3 &    55 &     3 & 638 &    SEKBO & likely Sgr\\
316.12666 &    1.11801 &  19.2 &    53 &     2 & 382 &   CRTS & possible Sgr?\\
331.79452 &    0.24400 &  19.1 &    51 &     2 & 527 &  Stripe 82 & possible Pisces/Sgr?\\
344.55949 &    0.27140 &  19.6 &    64 &     2 & 523 &   Stripe 82 & possible Pisces/Sgr?\\
345.44403 &   -9.98595 &  19.3 &    55 &     2 & 415 &    SEKBO & possible Pisces/Sgr?\\
\hline 
148.68102 &   16.03975 &  19.9 &    76 &     2 & 257 &    CRTS & likely Gemini\\
150.01648 &   22.52547 &  19.9 &    74 &     2 & 360 &   CRTS & possible Gemini\\
200.89584 &   -5.36901 &  20.1 &    83 &     2 & 553 &    SEKBO & possible Gemini\\
  \hline
266.29432 &   36.87977 &  19.1 &    52 &     2 & 531 &     CRTS & unknown\\
354.56833 &   25.72140 &  19.3 &    57 &     2 & 165 &      CRTS & unknown
 \enddata

\tablecomments{Groups of 2 of more RR Lyrae stars identified with a 500 pc linking length, with d $>$ 50 kpc, and not spatially coincident with dense RR Lyrae structures.}
\tablenotetext{a}{If an RRL group fell right on the edge of a distance bin, its recorded distance here may differ slightly from its location in Figure~\ref{aitoff_plots}.}
\tablenotetext{b}{Sizes of groups of 2 RR Lyrae may exceed the linking length (500 pc) by a small amount due to our method of group detection: When converting angular separations to physical separations when group finding, we use the middle of the distance bin being searched.  When calculating the physical size of each group, we utilize the mean distance of stars belonging to the group.}
%Thus, groups that are more distant than the middle of the bin can have max separations larger than the linking length itself. Note: some groups are not on plot because they lie slightly closer than 50kpc.
%\tablenotetext{b}{Based on MW Stellar Stream Figure made by B. Pila D'ez; not confirmed}
\label{final_table}
\end{deluxetable*}

%\subsection{Detection of Known, Unbound Halo Structures}\label{sec_known_structures}

A visual examination of the RR Lyrae distributions in Figure~\ref{aitoff_plots} reveals several known halo overdensities. In the top and middle panels, the Sagittarius (Sgr) leading arm is clearly seen with 180$^{\circ}$ $\lesssim$ RA $\lesssim$ 240$^{\circ}$.  By comparison with \citet{torrealba15a} the diffuse overdensity with 150$^{\circ}$ $\lesssim$ RA $\lesssim$ 180$^{\circ}$ and 0$^{\circ}$ $\lesssim$ Dec $\lesssim$ 30$^{\circ}$ in the top and middle panels is also likely Sgr stream material.  The Gemini stream (a likely distant wrap of Sgr, \citealp{Drake13b,Belokurov14a}) is clearly visible in the bottom panel of Figure~\ref{aitoff_plots} at 100$^{\circ}$ $<$ RA $<$120$^{\circ}$ and Dec $\sim$ 20$^{\circ}$, with a possible extension to the stream-like feature seen with 180$^{\circ}$ $\lesssim$ RA $\lesssim$ 210$^{\circ}$ and slightly negative declination.   The Pisces overdensity is also seen in the bottom panel of Figure~\ref{aitoff_plots} near RA $\sim$ 350$^{\circ}$ and Dec $\sim$ 0$^{\circ}$.  See \citet{Sesar07}, \citet{Watkins09a}, and \citet{Sesar10a} for the discovery of and follow-up discussions about Pisces. 

\subsection{Relatively Isolated RR Lyrae Groups in Public Catalogs}\label{sec_new_groups}

We selected 18 groups of two or more RR Lyrae that are not embedded in high surface density halo structures, for further inspection. We didn't require ``relatively isolated" to be all that isolated when selecting these groups, in order to be fully inclusive of possibly interesting groups while knowingly including interloper field halo groups.  While 18 RR Lyrae groups is loosely consistent with expectations for the number of field RR Lyrae groups found in Section~\ref{sec_contam}, the observed and predicted numbers aren't exactly comparable because of the significant incompleteness at d $>$ 50 kpc in these RR Lyrae catalogs.  

The properties of these groups are presented in Table~\ref{final_table} and their positions are highlighted in Figure~\ref{aitoff_plots} as the colored stars with boxes around them.  Two of these 18 groups are associated with known MW dwarf galaxies: the Bo\"otes I dwarf and Sextans. These results are as expected, because no Milky Way dwarf with more than 1 known RR Lyrae star resides within any public RR Lyrae survey's effective survey volume.  Although RR Lyrae are cataloged beyond these effective distances, completeness rapidly decreases and may be variable with location.  Sextans was found as a group of 2 RRab stars and resides with $d$ = 86 kpc in the MLS footprint, a survey with a 50\% completeness distance of 75 kpc and a maximum detection distance of 125 kpc.  Bo\"otes I was found as a group of 4 RRab stars and resides at 66 kpc in the CSS footprint, a survey with a 50\% completeness distance of 30 kpc and a maximum detection distance of 80 kpc. It is therefore mildly surprising that 4 of its 7 of its RRab stars were detected as a group. Ursa Minor (77 kpc) and Draco (76 kpc) are the only other dwarfs known with more than 2 RR Lyrae stars that lie within the maximum detection distance of a public RR Lyrae survey (CSS), but they are only a couple of kpc within that maximum distance and effective RR Lyrae completeness at the locations of those two dwarfs could easily be 0\% at those distances.

14 of these 18 groups appear to be associated with known halo structures.  Nine of the highlighted groups with 50 kpc $<$ d $<$ 60 kpc are likely to be associated with Sgr debris.  We use Figure 16 of \citet{torrealba15a} as a guide to classify the four with 120$^{\circ}$ $\lesssim$ RA $\lesssim$ 180$^{\circ}$ and Dec $\gtrsim$ 0$^{\circ}$ as likely Sgr debris, the one at RA $\sim$ 240$^{\circ}$ and Dec $\sim$ -25$^{\circ}$ as likely Sgr debris., and the one at RA $\sim$ 180$^{\circ}$ and Dec $\sim$ 25$^{\circ}$ as possible Sgr debris. The three groups with 315$^{\circ}$ $\lesssim$ RA $\lesssim$ 330$^{\circ}$ and Dec $\sim$ 0$^{\circ}$ are more ambiguous, and may belong to bits of Sgr or Pisces (for example, see the Pisces extension visible in Figure 10 of \citealt{Sesar10a}).  Similarly both highlighted groups between 60 kpc $<$ d $<$ 70 kpc, could potentially be associated with Sgr debris, though the one with RA $\sim$ 345$^{\circ}$ could also be part of the Pisces overdensity.  Finally, all of the isolated groups at d $>$ 70 kpc (not including the Sextans dwarf) may be associated with the Gemini structure.

The two structures we classify as ``unknown" reside at RA $\sim$ 355$^{\circ}$  and Dec $\sim$ 26$^{\circ}$ and RA $\sim$ 240$^{\circ}$  and Dec $\sim$ 40$^{\circ}$. Although the former lies in the general direction of the Tri-And overdensity, it is 20 kpc more distant \citep[e.g.][]{Sheffield14a}.  Because these two groups are at d $<$ 60 kpc and reside close to other halo RR Lyrae, it is likely that they are field RR Lyrae stars.  However, it would be worth following-up those regions of sky with imaging and/or a more complete RR Lyrae catalog.

\section{Discussion and Conclusions}

In this paper, we provide a proof-of-concept that groups of two or more RR Lyrae stars can distinctly reveal MW dwarf galaxies at d $>$ 50 kpc, and at Galactic latitudes and dwarf galaxy surface brightnesses that are inaccessible with current detection methods.  To do this, we simulated the RR Lyrae distributions of 13,200 dwarfs, assuming numbers of RR Lyrae stars consistent with the RRab specific frequencies of observed dwarfs, and applied a simple FOF algorithm to select groups of two or more RR Lyrae with a 2d linking length of 500 pc.   

This approach can recover most dwarfs similar to those known that are more luminous than M$_{\rm V}$ = -3.2 mag (with a possible dip in detectability for M$_{\rm V}$ = -5 mag to -4 mag, owing to observed specific frequency trends).  Because RR Lyrae can be distinguished from foreground stars near the Galactic plane, RR Lyrae could recover ultra-faint dwarfs similar to those known, but at low Galactic latitude. This approach also recovers nearly all dwarfs with surface brightnesses of 31 mag arcsec$^{-2}$ and fainter that are more luminous than M$_{\rm V}$ = -6 mag, and will recover very low luminosity dwarfs (-4.5 $\lesssim$ M$_{\rm V}$ $\lesssim$ -3.2) with high RR Lyrae specific frequency to similarly low surface brightnesses.  Groups of RR Lyrae may thus provide the most effective means to test the predicted existence of stealth \citep{bullock10a} and fossil galaxies \citep{Bovill11a}.

We used the simulated stellar halo catalogs of \citet{Lowing15} to demonstrate that only modest contamination from groups of two or more field halo RR Lyrae is expected at d $>$ 50 kpc, when excluding groups near major field halo structures.  In particular we found  (i) FOF searches with a 500 pc linking length are appropriate to find dwarf galaxies with d $>$ 100 kpc as groups of 2 more more RR Lyrae, and (ii) groups with as few as 2 RR Lyrae can also distinctly identify dwarf galaxies with 50 kpc $<$ d $<$ 100 kpc, however field contamination should be limited at these nearer distances by more strictly excluding groups of only 2 RR Lyrae near larger-scale spatial clusterings of RR Lyrae.  

We searched several public RR Lyrae catalogs (CRTS - CSS and MLS, LSQ, SEKBO, and Stripe 82) for relatively isolated groups of RR Lyrae to identify possible dwarf galaxy candidates with 50 kpc $<$ d $<$ 100 kpc.  However, because these catalogs are very incomplete to RR Lyrae at d $>$ 50 kpc, we can't use this search to put quantitative limits on low surface brightness dwarfs.  The Bo\"otes I (66 kpc) and Sextans (86 kpc) dwarfs were found as groups of 4 and 2 RR Lyrae in the CRTS and LSQ catalogs, respectively, as expected given survey depths.  This is far fewer than their known number of RR Lyrae because the CRTS and LSQ catalogs are severely incomplete at those distances.  We associated a number of other relatively isolated groups of only 2 RR Lyrae with known halo structures, and identified two interesting groups that are not obviously associated with known structures.

One of the most exciting regions of dwarf galaxy discovery space that RR Lyrae may open up is Galactic latitudes closer than $\sim$25$^{\circ}$ to the plane, because the detectability of dwarfs with traditional resolved star counts drops off rapidly at latitudes with $b \lesssim$ 20$^{\circ}$ \citep{Walsh09}.  We caution that the simplified proof-of-concept calculations presented in this paper do not include the detection limiting effects of Galactic extinction, the increasing uncertainties in extinction (and therefore of RR Lyrae distance) with decreasing Galactic latitude, and the possible contamination of thick disk stars in RR Lyrae group finding at very low Galactic latitudes.  Qualitatively, Galactic extinction will decrease the effective RR Lyrae survey volume of each survey (especially at $b \lesssim$ 20$^{\circ}$) and distance uncertainties will soften the detectability of dwarfs.  Even detection limits that are $>$2 magnitudes shallower at the lowest Galactic latitudes, would represent enormous progress; More detailed simulations can explore these possibilities in the future.  Including RR Lyrae periods in RR Lyrae group finding will help mitigate the impact of distance uncertainties and thick disk contamination on finding ultra-faint dwarfs at very low latitudes.  Such dwarfs are populated by particularly long period RR Lyrae \citep[e.g.][]{Catelan09,Boettcher13}.
 
PanSTARRS 1 may be the most exciting existing dataset for discovering new MW dwarfs as groups of RR Lyrae stars.  Its RR Lyrae survey volume could include $\sim$15\% of the MW's dwarf galaxies (Section~\ref{sec_elvis}), and it includes a large area at relatively low Galactic latitudes (unlike P/i/ZTF).  If PanSTARRS 1 data produce an RR Lyrae catalog that is reasonably complete to 100 kpc in areas of relatively low Galactic extinction, then groups of RR Lyrae at low latitude (or the lack thereof) could test the apparent significant spatial anisotropy in MW dwarf galaxies.  The PS1 RR Lyrae catalog should also reveal candidates for extremely low surface brightness MW dwarf galaxies, if they do exist.

Looking ahead, complimenting traditional search techniques with RR Lyrae in LSST may be the only way to obtain the most complete possible census of Milky Way dwarf galaxies - through the Galactic plane and to surface brightnesses of 31 mag arcsec$^{-2}$ and fainter.  LSST is the only planned survey that will be both wide-field and deep enough to grasp the majority of the MW's halo with RR Lyrae stars.  RR Lyrae identified in LSST's wide-fast-deep survey will include $\sim$45\% of the MW's predicted dwarf galaxy companions (even when accounting for Galactic extinction).  This quoted volume to be covered by the wide-fast-deep survey is a lower limit to LSST's RR Lyrae survey volume, because $\sim$10\% of survey time will be dedicated to thousands of additional square degrees of the Northern Ecliptic Spur, the Galactic plane, the Magellanic Clouds, and other special fields.  RR Lyrae in the LSST footprint will be efficiently identified to d $\sim$ 600 kpc, enabling RR Lyrae-based maps of stellar structures and dwarf galaxies to well beyond the edge of the MW.  

\acknowledgments

BW and MB acknowledge support from NSF AST-1151462.  We thank J. Cohen, J. Hargis, N. Hernitschek, D. Sand, B. Sesar, J. Strader, J. Surace, and J. VanderPlas for helpful conversations over the course of this paper writing and B. Sesar, B. Laevens, and N. Martin for inspecting an early dwarf galaxy candidate in PanSTARRS 1 data, which revealed an error in our initial implementation.  We thank Wenting Wang and Carlos Frenk for sharing their simulations and answering our questions about their simulated star catalogs, and thank Wenting Wang and Andrew Cooper for providing the luminosities and stellar masses assigned to the sub-halos in \citet{Lowing15}.  We thank Joe Cammisa at Haverford College for his computing support.  Last, but not least, thank you to the Aspen Center for Physics and the NSF Grant \#1066293 for hospitality during the writing and editing of this paper.  This work has made use of NASA's Astrophysics Data System.

\bibliographystyle{apj}
%\bibliography{RRL}

\end{document}